\documentclass[twocolumn,PRL,showpacs,nofootinbib,amsmath,amssymb]{revtex4-1}

\usepackage{newtxtext,newtxmath}
\usepackage[T1]{fontenc}
\usepackage{ae,aecompl}
\usepackage{graphicx}	% Including figure files
\usepackage{amsmath}	% Advanced maths commands
\usepackage{amssymb}	% Extra maths symbols
\usepackage[usenames,dvipsnames,svgnames,table]{xcolor}
\usepackage{subfig}
\usepackage{ulem}
\usepackage{float}
\usepackage{multirow}
\usepackage{hyperref}

\definecolor{darkred}{RGB}{175,0,0}

\begin{document}

\title{Lyman-$\alpha$ forest constraints on Primordial Black Holes as Dark Matter}

\begin{flushright}
CERN-TH-2019-029
\end{flushright}
\vfil

\author{Riccardo Murgia,$^{1,2,3}$, Giulio Scelfo,$^{1,2,3}$, Matteo Viel,$^{1,2,3,4}$, and Alvise Raccanelli$^{5}$}
\affiliation{
$^{1}$SISSA, Via Bonomea 265, 34136 Trieste, Italy\\
$^{2}$INFN, Sezione di Trieste, via Valerio 2, 34127 Trieste, Italy\\
$^{3}$IFPU, Institute for Fundamental Physics of the Universe, via Beirut 2, 34151, Trieste, Italy\\
$^{4}$INAF/OATS, Osservatorio Astronomico di Trieste, via Tiepolo 11, I-34143 Trieste, Italy\\
$^{5}$Theoretical Physics Department, CERN, 1 Esplanade des Particules, CH-1211 Geneva 23, Switzerland
}

\date{\today}

\begin{abstract}
The renewed interest in the possibility that primordial black holes (PBHs) may constitute a significant part of the dark matter has motivated revisiting old observational constraints, as well as developing new ones.
We present new limits on the PBH abundance, from a comprehensive analysis of high-resolution, high-redshift Lyman-$\alpha$ forest data.
Poisson fluctuations in the PBH number density induce a small-scale power enhancement which departs from the standard cold dark matter prediction. 
Using a grid of hydrodynamic simulations exploring different values of astrophysical parameters, {we obtain a marginalized upper limit on the PBH mass of $f_{\rm PBH}M_{\rm PBH} \sim 60~M_{\odot}$ at $2\sigma$, when a Gaussian prior on the reionization redshift is imposed, preventing its posterior distribution to peak on very high values, which are disfavoured by the most recent estimates obtained both through Cosmic Microwave Background and Inter-Galactic Medium observations. Such bound weakens to $f_{\rm PBH}M_{\rm PBH} \sim 170~M_{\odot}$, when a conservative flat prior is instead assumed. Both limits significantly improves previous constraints from the same physical observable.} We also extend our predictions to non-monochromatic PBH mass distributions, ruling out large regions of the parameter space for some of the most viable PBH extended mass functions.
\end{abstract}

\maketitle

%%%%%%%%%%%%%%%%% BODY OF PAPER %%%%%%%%%%%%%%%%%%

{\it Introduction.}
\label{sec:intro}
Primordial Black Holes (PBHs) were first theorized decades ago~\cite{Hawking:pbh1971}. Many proposals were made for their formation mechanism, such as collapsing large fluctuations produced during inflation~\cite{ivanov:pbhfrominflationI, Bellido:pbh, ivanov:pbhfrominflationII}, collapsing cosmic string loops~\cite{Polnarev88, HAWKING1989237, WICHOSKI1998191}, domain walls~\cite{BEREZIN198391, Ipser84}, bubble collisions~\cite{Crawford82, LA1989375}, or collapse of exotic Dark Matter (DM) clumps~\cite{shandera:pbhfromdarkmatter}.

After the first Gravitational Wave (GW) detection revealed merging Black Hole (BH) binaries of masses $\mathcal{O}(10 \: M_{\odot})$~\cite{abbott:firstligodetection, abbott:firstligodetectionproperties}, the interest toward PBHs as DM candidates has revived~\cite{bird:pbhasdarkmatter}. Several proposals to determine the nature of the merging BH progenitors have been made, involving methods as GW$\times$LSS cross-correlations~\cite{raccanelli:pbhprogenitors, Scelfo_2018}, BH binary eccentricities~\cite{cholis:orbitaleccentricities}, BH mass function studies~\cite{kovetz:pbhmassfunction, kovetz:pbhandgw}, lensing of fast radio bursts~\cite{munoz:fastradioburst}.

Several constraints on the PBH abundance have been determined through different observables, such as gravitational lensing~\cite{barnacka:femtolensingconstraint, katz:femtolensingconstraint, griest:keplerconstraint, niikura:microlensingconstraint, tisserand:microlensingconstraint, calchinovati:microlensingconstraint, alcock:microlensingconstraint, mediavilla:microlensingconstraint, wilkinson:millilensingconstraint, zumalacarregui:supernovaconstraint}, dynamical~\cite{graham:whitedwarfconstraint, capela:neutronstarconstaint, quinn:widebinaryconstraint, brandt:ufdgconstraint, Raidal:2017mfl, alihaimoud:pbhmergerrate, Raidal:2018bbj, magee:mergerrate}, and accretion effects~\cite{gaggero:accretionconstraints, ricotti:cmbconstraint, alihamoud:pbhaccretion, poulin:cmbconstraint, bernal:cmbconstraint}. 
Nevertheless, varying the numerous assumptions involved might significantly alter these limits~\cite{Aloni2017,bellomo:emdconstraints,Nakama:NG}, making the investigation towards PBHs as DM candidates still fully open. Specifically, two mass regimes are currently of large interest: $\mathcal{O}(10^{-10} M_{\odot})$, and $\mathcal{O}(10 \: M_{\odot})$~(see~\cite{Sasaki_2018, Carr_Silk_2018, Ali-Haimoud:2019khd} for details).

A mostly unexplored method for constraining the PBH abundance is offered by the Lyman-$\alpha$ forest, which is the main manifestation of the Inter-Galactic Medium (IGM), and represents a powerful tool for tracing the DM distribution at (sub-) galactic scales~(see,~e.g.,~\cite{Viel:2001hd,Viel2005,Viel:2013apy}).
Lyman-$\alpha$ data were used about 15 years ago to set an upper limit of few~$10^4 M_{\odot}$ on PBH masses, assuming all DM made by PBHs with the same mass~\cite{Afshordi:2003zb}.
In this work, we update and improve such limit, using the highest resolution Lyman-$\alpha$ forest data up-to-date~\cite{Viel:2013apy}, and a new set of high-resolution hydrodynamic simulations. %that allow a more precise modeling of the full 1D flux power.
Furthermore, we generalize our results to different PBH abundances and non-monochromatic mass distributions. \\ \vspace{-0.25cm}

%This Letter is organized as follows: in Section~\ref{sec:pbh} we discuss the impact on the matter power spectrum due to the existence of PBHs; in Section~\ref{sec:EMD} we extend the discussion to non-monochromatic PBH mass distributions; in Section~\ref{sec:method} we present the data set and the methods that we adopted for our analyses; in Section~\ref{sec:results} we report and discuss the results that we have obtained; in Section~\ref{sec:concl} we draw our conclusions. \\

{\it Poisson noise and impact on the matter power spectrum}.
Stellar-mass PBHs would cause observable effects on the matter power spectrum; due to discreetness, a small-scale plateau in the linear power spectrum is induced by a Poisson noise contribution~\cite{meszaros:primeval, Afshordi:2003zb, Carr_Silk_2018, Gong:2017sie}.

If PBHs are characterized by a Monochromatic Mass Distribution (MMD), they are parameterized by their mass, $M_{\mathrm{PBH}}$, and abundance, so that the fraction parameter $f_{\mathrm{PBH}} \equiv \Omega_{\rm PBH}/\Omega_{\rm DM} = 1$ where all DM is made of PBHs.

If PBHs are randomly distributed, their number follows a Poisson distribution, and each wavenumber $k$ is associated to an overdensity $\delta_{\mathrm{PBH}}(k)$, due to Poisson noise. The PBH contribution to the power spectrum is thus defined as
\begin{equation}
\label{eq:p_pbh}
P_{\mathrm{PBH}}(k) = \langle | \delta_{\mathrm{PBH}} (k)|^2  \rangle = \frac{1}{n_{\mathrm{PBH}}} \, ,
\end{equation}
where $n_{\mathrm{PBH}}$ is the comoving PBH number density
,~i.e.
\begin{equation}\label{eq:number_density}
 n_{\mathrm{PBH}} = \dfrac{\Omega_{\mathrm{DM}} \rho_{\mathrm{cr}} f_{\mathrm{PBH}}}{M_{\mathrm{PBH}}},
\end{equation}
with $\rho_{\mathrm{cr}}$ being the critical density of the universe.
%From Equation~\eqref{eq:number_density} it is immediate to see a degeneracy between $f_{\mathrm{PBH}}$ and $M_{\mathrm{PBH}}$: different combinations of PBH mass and abundance correspond to the same number density if the fraction $f_{\mathrm{PBH}}/M_{\mathrm{PBH}}$ is the same. 
Since $n_{\rm PBH}$ is a $k$-independent quantity, $P_{\rm PBH}$ is scale-invariant.
%In other words, the existence of PBHs induces a small-scale plateau departing from the standard $\Lambda$CDM prediction. 

One can interpret the PBH overdensity as an isocurvature perturbation~\cite{Afshordi:2003zb, Gong:2017sie}. Hence, the total power spectrum can be written as:
\begin{equation}\label{eq:CDM_ps}
P_{\mathrm{CDM}}(k,z) = D^2(z)\left(T^2_{\mathrm{ad}}(k) P_{\mathrm{ad}}+ T^2_{\mathrm{iso}}(k) P_{\mathrm{iso}}\right) \, ,
\end{equation}
where $D(z)$ is the growth factor, $P_{\rm iso}$ is the isocurvature power spectrum, and $P_{\mathrm{ad}} \propto A_s k^{n_s}$ is the primordial adiabatic power spectrum. $T_{\mathrm{ad}}$ and $T_{\mathrm{iso}}$ are the adiabatic and isocurvature transfer functions, respectively.
The PBH linear power spectrum is thus defined by: 
\begin{equation}
P _ { \mathrm { iso} } = {f^2_{\rm PBH}} P _ { \mathrm { PBH} } = \frac{2\pi^2}{k^3} A _ { \mathrm { iso } } \left( \frac { k } { k _ { * } } \right) ^ { n _ { \mathrm { iso } } - 1 } \, ,
\end{equation}
where we set the pivot scale $k_* = 0.05$/Mpc, and the primordial isocurvature tilt $n _ { \mathrm { iso }} =4$ in order to ensure the scale-invariance. Since the adiabatic power spectrum evolves as $k^{-3}$ at large $k$, the isocurvature contribution is expected to become important only at the scales probed by Lyman-$\alpha$ forest; $A_{\mathrm{iso}}$ sets the amplitude of the isocurvature modes, depending on the PBH mass considered; we can then express the isocurvature-to-adiabatic amplitude ratio:
\begin{equation}\label{eq:f_iso}
f_{\mathrm{iso}}=\sqrt{\frac{A_{\mathrm{iso}}}{A_s}}= \sqrt { \frac { k _ { * } ^ { 3 } f _ { \mathrm { PBH } }^2 } { 2 \pi ^ { 2 } n_{ \mathrm { PBH }}} \frac { 1 } { A _ { \mathrm { s } } } } = \sqrt { \frac { k _ { * } ^ { 3 } M _ { \mathrm { PBH } } f _ { \mathrm { PBH } } } { 2 \pi ^ { 2 } \Omega _ { \mathrm { CDM } } \rho _ { \mathrm { cr } } } \frac { 1 } { A _ { \mathrm { s } } } } \, ,
\end{equation}
where the last equality holds only for MMDs. Different combinations of PBH mass and abundance correspond to the same isocurvature-to-adiabatic amplitude ratio if the quantity $f_{\mathrm{PBH}} M_{\mathrm{PBH}}$ is the same.
\begin{figure}
   \centering
   \includegraphics[width=1.0\linewidth]{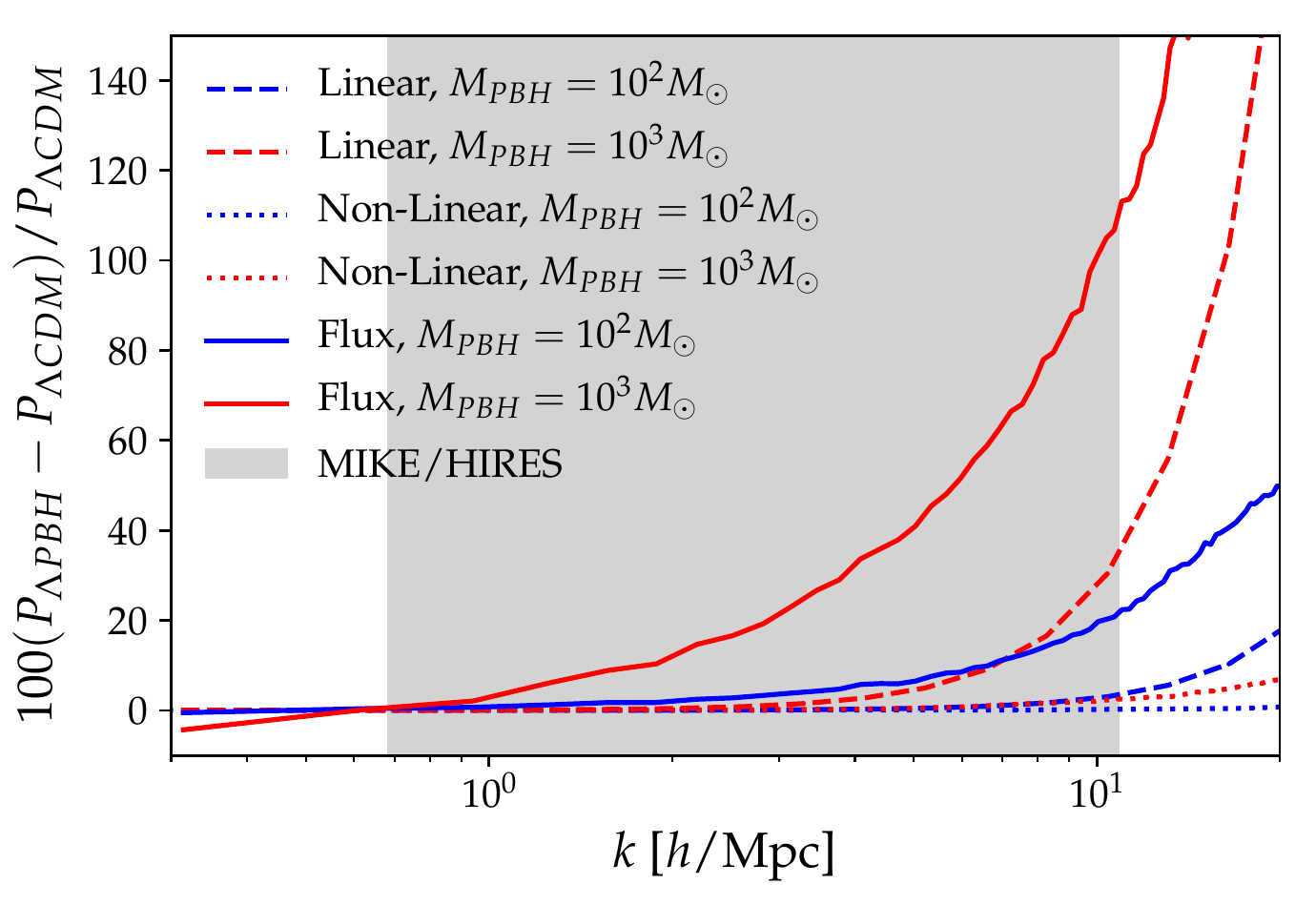}
   \caption{Relative difference, at redshift $z=5$, between $\Lambda \mathrm{CDM}$ and $\Lambda \mathrm{PBH}$ scenarios for the linear matter ($dashed$), non-linear matter ($dotted$) and 1D flux ($solid$) power spectra. Blue/red colors correspond to $M_{\mathrm{PBH}}=\{10^2,10^3\}~M_{\odot}$, respectively, for a monochromatic scenario with $f_{\mathrm{PBH}} = 1$. The gray shaded area refers to the scales covered by Lyman-$\alpha$ forest data.}
   \label{fig:ps_class}
\end{figure}
In our framework, the effect on the linear matter power spectrum due to the presence of isocurvature modes consists of a power enhancement with respect to the standard $\Lambda$CDM spectrum, in the form of a small-scale plateau.
%As a straightforward consequence of Equation~\eqref{eq:p_pbh}, the larger the PBH mass, the stronger the effect.

\begin{figure}
	\centering
	\includegraphics[width=0.95\linewidth]{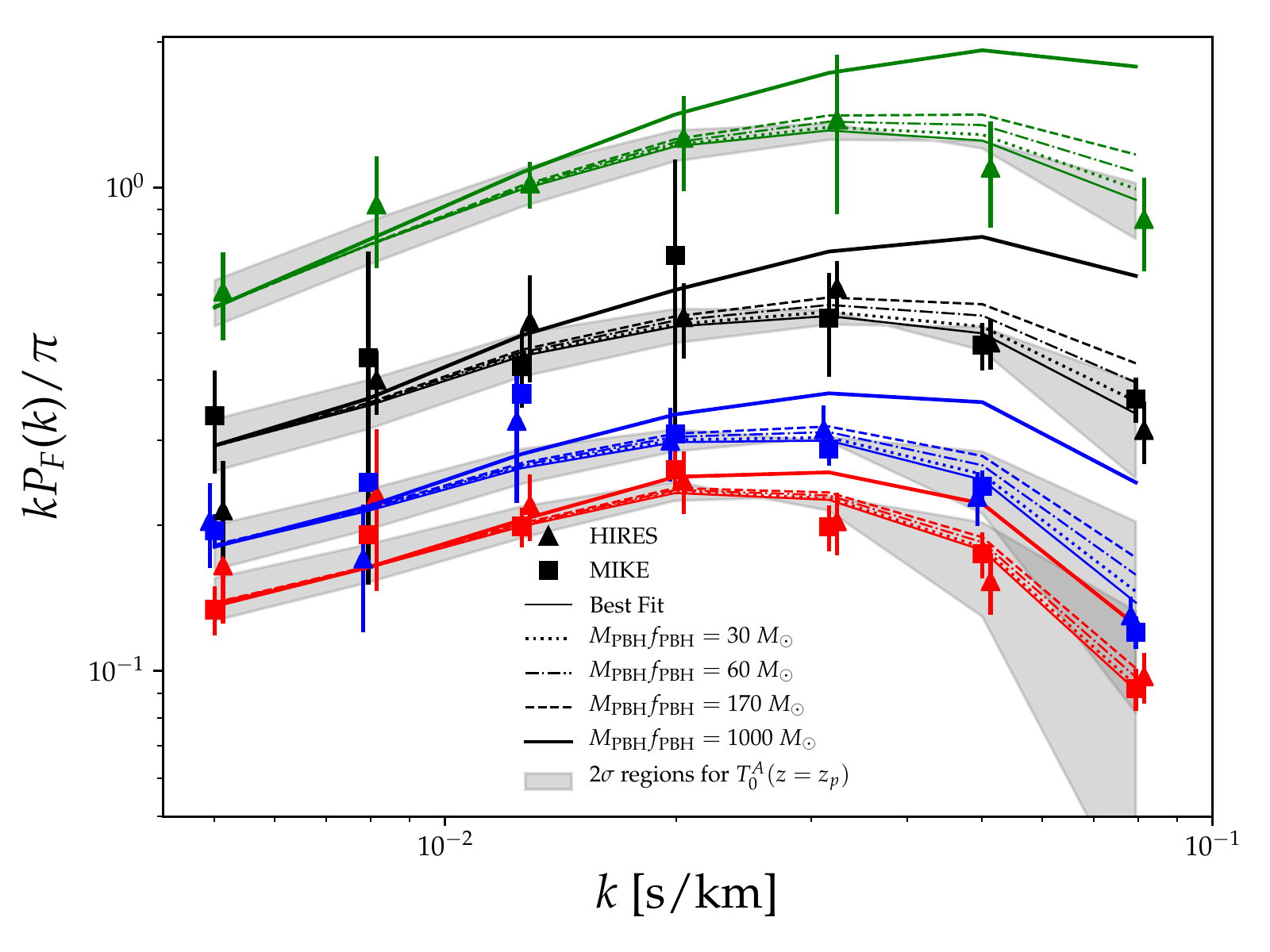}
	\caption{1D flux spectra for $\Lambda$CDM and $\Lambda$PBH, for different PBH masses. Symbols are data from MIKE/HIRES, lines are obtained by interpolating in the $(M_{\rm PBH}f_{\rm PBH})$-space defined by our simulations; while the best fit is technically for $M_{\rm PBH}\neq 0$, it is indistinguishable from the $\Lambda$CDM case.
		Red, blue, black and green indicate $z={4.2, 4.6, 5.0, 5.4}$, respectively. {The grey dashed areas represent regions sampled by flux power spectra corresponding to values for $T_0^A$ spanning its marginalized 2$\sigma$ interval.}}
	\label{fig:ps_flux}
\end{figure}

In Figure~\ref{fig:ps_class} we provide the relative differences with respect to a pure $\Lambda$CDM scenario for the 3D linear and non-linear matter power spectra, at redshift $z=5$, for $\Lambda$PBH models with $M_{\rm PBH} = \{10^2, 10^3\} M_{\odot}$, assuming $f_{\rm PBH} = 1$. We also show the 1D flux power spectra, which are the Lyman-$\alpha$ forest observables, associated to the same $\Lambda \mathrm{PBH}$ models. The gray shaded area refers to the scales covered by our Lyman-$\alpha$ data set, obtained from MIKE/HIRES spectrographs. 
The non-linear power spectra have been extracted from the snapshots of cosmological simulations, thus they include both the linear contribution (encoded in the initial conditions) and, on top of that, the effects of the non-linear evolution computed by the numerical simulation itself. The PBH contribution is thereby included in the initial conditions, whereas during the non-linear evolution both the isocurvature and adiabatic DM modes are treated as cold and collisionless (see \textit{Data set and methods} Section for further details). 
It can be easily seen how non-linearities in the 3D matter power spectrum wash out the differences induced by the presence of PBHs. On the other hand, the 1D flux spectra are a much more effective observable to probe the small-scale power. \\
{The Lyman-$\alpha$ forest observable is the 1D flux power spectrum, which, being a projection of 3D non-linear matter power spectrum, is
an ideal tracer for the small-scale DM distribution along our lines of sight.
In Figure~\ref{fig:ps_flux} we show the 1D flux power for the $\Lambda$CDM model, together with the spectra corresponding to different values of $M_{\mathrm{PBH}}$. Symbols refer to MIKE/HIRES data. To exhibit the variations in the flux power induced by different IGM thermal histories, we also show, as grey dashed areas, the impact of different IGM temperature evolutions.}\\
% We use 1D flux spectra as non-linearities in the 3D matter power spectrum wash out the differences induced by the presence of PBHs and they introduce modeling complications. \\

\vspace{-0.25cm}

{\it Extended Mass Distributions}.
\label{sec:EMD}
The PBH formation is, in the most standard case, due to large perturbations in the primordial power spectrum; while the exact details of the peak required to form PBH and how this is linked to the real-space overdensities are still unclear~\cite{Kalaja}, PBHs could have an extended mass function. Moreover, a non-monochromatic mass distribution would be created by different merger and accretion history of each PBH.
% ; the high-mass tail of the distribution may account for the seeds of supermassive BHs~\cite{Bernal:2017nec, Carr_Silk_2018}.
General methods to convert MMD constraints to limits on Extended Mass Distributions (EMDs) have been developed in~\cite{Carr:2017jsz,bellomo:emdconstraints}.

%It is known that the size of a given PBH is of the same order of the particle horizon at its formation time \AR{assumed and not fully true}.
%Since PBHs can form at different times in the early universe, their corresponding number density $n_{\mathrm{PBH}}$ evolves while their formation is taking place \AR{?}. Conversely, when PBHs stop forming, $n_{\mathrm{PBH}}$ is fixed\AR{?}. 
%Under this assumption, 
The extension to EMDs of the observable considered here arises naturally from the second equality in Equation~\eqref{eq:f_iso}, by directly taking the PBH number density corresponding to a given EMD.
Consider EMDs in the form:
\begin{equation}
    \dfrac{d n_{\mathrm{PBH}}}{d\ln M_{\rm PBH}} = f_{\mathrm{PBH}}\rho_{\mathrm{DM}} \dfrac{d \Phi_{\mathrm{PBH}}}{dM_{\rm PBH}} \, ,
\end{equation}
where the function $d \Phi_{\mathrm{PBH}}/{dM_{\rm PBH}}$ describes the EMD shape, and $\rho_{\mathrm{DM}} = \Omega_{\mathrm{DM}} \rho_{\mathrm{cr}}$. Given an EMD, one can define the so-called Equivalent Mass $M_{\mathrm{eq}}$, which is the mass of a MMD providing the same observational effect.\\
The conversion is given by:
\begin{equation}
    f_{\mathrm{PBH}}^2 \Biggl[\dfrac{\Omega_{\mathrm{DM}} \rho_{\mathrm{cr}} f_{\mathrm{PBH}}}{M_{\mathrm{eq}}}\Biggl]^{-1} = \dfrac{f_{\mathrm{PBH}}^2}{n_{\mathrm{PBH}}} = f_{\mathrm{PBH}}^2 \Biggl[\int \dfrac{d n_{\mathrm{PBH}}}{dM_{\rm PBH}} dM_{\rm PBH}\Biggl]^{-1}
\end{equation}
where we assume that the PBH abundances are the same for both the MMD and EMD cases. We finally have:
\begin{equation}\label{eq:eq_mass}
   M_{\mathrm{eq}} = \Biggl[ \int \dfrac{1}{M_{\rm PBH}} \dfrac{d\Phi}{dM_{\rm PBH}} dM_{\rm PBH}\Biggl]^{-1}  \, .
\end{equation}
We consider two popular EMDs: \textit{Lognormal} and \textit{Powerlaw}.

The \textit{Lognormal} EMD~\cite{dolgov_silk} is defined by
\begin{equation}
\frac{d\Phi_\mathrm{PBH}}{dM_{\rm PBH}} = \frac{\exp \left\{ {-\frac{\ln^2(M_{\rm PBH}/\mu) }{2\sigma^2}} \right\} } {\sqrt{2\pi}\sigma M_{\rm PBH}} \, ,
\end{equation}
where $\sigma$ and $\mu$ are the standard deviation and mean of the PBH mass, respectively. Such function describes,~e.g.,~the scenario of PBHs forming from a smooth symmetric peak in the inflationary power spectrum~\cite{green:2016,kannike:2017}.

The \textit{Powerlaw} EMD, corresponding to PBHs formed from collapsing cosmic strings or scale-invariant density fluctuations~\cite{carr:1975primordialmassspectrum}, is given by
\begin{equation}
\frac{d\Phi_\mathrm{PBH}}{dM_{\rm PBH}} = \frac{\mathcal{N}_{\rm PL}}{M_{\rm PBH}^{1-\tilde{\gamma}}}\Theta(M_{\rm PBH}-M_\mathrm{min})\Theta(M_\mathrm{max}-M_{\rm PBH}) \, ,
\end{equation}
characterized by an exponent $\tilde{\gamma}\in(-1,+1)$, a mass interval $(M_\mathrm{min}, M_\mathrm{max})$, and a normalization factor $\mathcal{N}_{\rm PL}$; $\Theta$ is the Heaviside step function. \\ \vspace{-0.25cm}
%For further details on any of the EMDs considered in this work, we address the reader to the aforementioned references.

{\it Dataset and methods}.
\label{sec:method}
To extract limits on the PBH abundance from the Lyman-$\alpha$ forest, we adapted the method proposed in~\cite{Murgia:2018now}. 
% for testing suppressed linear matter power spectra with structure formation data. 
We built a new grid of hydrodynamic simulations in terms of the properties of PBHs, corresponding to initial linear power spectra featuring a small-scale plateau. Beside that, our analyses rely on a pre-computed multidimensional grid of hydrodynamic simulations, associated to several values of the astrophysical and cosmological parameters affecting the Lyman-$\alpha$ flux power spectrum.
% , sampling all the viable volume of the corresponding parameter space. 
Simulations have been performed with {\texttt{GADGET-III}}, a modified version of the public code {\texttt{GADGET-II}}~\cite{Springel:2000yr,Springel:2005mi}. Initial conditions have been produced with {\texttt{2LPTic}}~\cite{Crocce:2006ve}, at $z=199$, with input linear power spectra for the $\Lambda$PBH models obtained by turning on the isocurvature mode in {\texttt{CLASS}}~\cite{Blas:2011rf}.

Our reference model simulation~\cite{Murgia:2018now, Irsic:2017ixq} has a box length of $20/h$ comoving Mpc with $2 \times 768^3$ gas and CDM particles in a flat $\Lambda$CDM universe
with cosmological parameters as in~\cite{Ade:2015xua}.
%: $\Omega_m = 0.301$, $\Omega_b = 0.0457$, $n_s = 0.961$, $H_0 = 70.2$~km~s$^{-1}$~Mpc$^{-1}$, $\sigma_8 = 0.829$, and $z_{\rm reio} = 9$. 

For the cosmological parameters to be varied, we sample different values of $\sigma_8$,~i.e.,~the normalization of the linear power spectrum, and $n_{\rm eff}$, the slope of the power spectrum evaluated at the scale probed by the Lyman-$\alpha$ forest ($k_{\alpha} = 0.009$ s/km)~\cite{seljak2006,McDonald:2004eu,Arinyo-i-Prats:2015vqa}.
We included five different simulations for both $\sigma_8$ ($[0.754, 0.904]$) and $n_{\rm eff}$ ($[-2.3474,-2.2674]$). 
Additionally, we included simulations corresponding to different values for the instantaneous reionization redshift,~i.e.,~$z_{\rm reio} = \{7,9,15\}$.

Regarding the astrophysical parameters, we modeled the IGM thermal history with amplitude $T_0$ and slope ${\gamma}$ of its temperature-density
relation, parameterized as $T=T_0(1+\delta_\mathrm{IGM})^{{\gamma}-1}$, with $\delta_{\mathrm{IGM}}$ being the IGM overdensity~\cite{hui97}.
We use simulations with temperatures at mean density $T_0(z = 4.2) = \{6000, 9200, 12600\}$~K,
evolving with redshift, and a set of three values for the slope of the temperature-density relation, ${\gamma}(z = 4.2) = \{0.88, 1.24, 1.47\}$.
The redshift evolution of both $T_0$ and ${\gamma}$ are parameterized as power laws, such that $T_0(z) = T_0^A[(1+z)/(1+z_p)]^{T_0^S}$ and ${\gamma}(z) = {\gamma}^A[(1+z)/(1+z_p)]^{{\gamma}^S}$, where the pivot redshift $z_p$ is the redshift at which most of the Lyman-$\alpha$ forest pixels are coming from ($z_p = 4.5$).
The reference thermal history is defined by $T_0(z = 4.2) = 9200$ and ${\gamma}(z = 4.2) = 1.47$~\cite{bolton17}.

Furthermore, we considered the effect of ultraviolet (UV) fluctuations of the ionizing background, controlled by the parameter $f_{\rm UV}$. Its template is built from three simulations with $f_{\rm UV} = \{0, 0.5, 1\}$, where $f_{\rm UV} = 0$ corresponds to a spatially uniform UV background~\cite{Irsic:2017ixq}.
We also included 9 grid points obtained by rescaling the mean Lyman-$\alpha$ flux $\bar{F}(z)$, namely $\{0.6,0.7,0.8,0.9,1.0,1.1,1.2,1.3,1.4\} \times \bar{F}_{\rm REF}$, with reference values given by SDSS-III/BOSS measurements~\cite{boss2013}.
% With the goal to have a more refined grid in terms of mean fluxes, 
We also considered 8 additional values, obtained by rescaling the optical depth $\tau = -\ln\bar{F}$,~i.e.~$\{0.6,0.7,0.8,0.9,1.1,1.2,1.3,1.4\} \times \tau_{\rm REF}$.

Concerning the PBH properties, we extracted the flux power spectra from 12 hydrodynamic simulations ($512^3$ particles; 20 comoving Mpc/$h$ box length) corresponding to the following PBH mass and fraction products: $\log(M_{\rm{PBH}}f_{\rm PBH})=\{1.0, 1.5, 2.0, 2.2, 2.3, 2.4, 2.5, 2.6, 2.7, 3.0, 3.5, 4.0\}$.
For this set of simulations, astrophysical and cosmological parameters have been fixed to their reference values, and the equivalent $\Lambda$CDM flux power was also determined.

We use an advanced interpolation method,~the \textit{Ordinary Kriging} method~\cite{webster2007geostatistics}, particularly suitable to deal with the sparse, non-regular grid defined by our simulations. 
{Such method basically consists in predicting the value of the flux power at a given point by computing a weighted average of all its known values, with weights inversely proportional to the distance from the considered point.}
The interpolation is in terms of ratios between the flux power spectra of the $\Lambda$PBH models and the reference $\Lambda$CDM one. We first interpolate in the astrophysical and cosmological parameter space for the $\Lambda$CDM case, then correct all the $(M_{\rm PBH}f_{\rm PBH})$-grid points accordingly, and finally interpolate in the $(M_{\rm PBH}f_{\rm PBH})$-space.
This procedure relies on the assumption that the corrections due to non-reference astrophysical or cosmological parameters are universal, so that we can apply the same corrections computed for the $\Lambda$CDM case to the $\Lambda$PBH models as well.

Our datasets are the MIKE and HIRES/KECK samples of quasar spectra, at $z=\{4.2,4.6,5.0,5.4\}$, in 10 $k$-bins in the range $[0.001-0.08]$~s/km, with spectral resolution of 13.6 and 6.7~s/km~\cite{Viel:2013apy}.
%As in the analyses of~\cite{Murgia:2018now, Irsic:2017ixq}, 
We consider only measurements at $k > 0.005$ s/km, to avoid systematic uncertainties due to continuum fitting. 
Moreover, we did not use MIKE highest redshift bin.~\cite{Viel:2013apy}.
% , for which the errors on the flux power spectra are very large. 
We thus have a total of 49 $(k,z)$ data-points.\\ 
\begin{table}
	\setlength{\tabcolsep}{1pt}\renewcommand{\arraystretch}{1}
	\centering
	\normalsize{
		\begin{tabular}{|c|cc|cc|} \hline
			& \multicolumn{2}{c|}{\scriptsize\bf Flat prior on $z_{\rm reio}$} & \multicolumn{2}{c|}{\scriptsize\bf Gaussian prior on $z_{\rm reio}$} \\
			\hline \scriptsize\bf{Parameter}       & \scriptsize\bf{(2$\sigma$)} & \scriptsize\bf{Best Fit} & \scriptsize\bf{(2$\sigma$)} & \scriptsize\bf{Best Fit} \\ \hline
			$\bar{F}(z=4.2)$               & [0.35,~0.41]        &  0.37            & [0.35,~0.41]        &  0.37            \\  
			$\bar{F}(z=4.6)$               & [0.26,~0.34]        &  0.28            & [0.27,~0.34]        &  0.28            \\  
			$\bar{F}(z=5.0)$               & [0.15,~0.25]        &  0.20            & [0.15,~0.23]        &  0.16            \\  
			$\bar{F}(z=5.4)$               & [0.03,~0.12]        &  0.08            & [0.04,~0.11]        &  0.05            \\  
			$T_0^A~[10^{4}$~K]             & [0.44,~1.36]        &  0.72            & [0.46,~1.44]        &  0.84            \\  
			$T_0^S$ 	               & [-5.00,~3.34]       &  -4.47           & [-5.00,~3.35]       &  -4.53           \\ 
			${\gamma}^A$                   & [1.21,~1.60]        &  1.51            & [1.19,~1.61]        &  1.44            \\  
			${\gamma}^S$                   & [-2.43,~1.30]       &  -1.76           & [-2.25,~1.51]       &  0.46            \\ 
			$\sigma_8$                     & [0.72,~0.91]        &  0.79            & [0.72,~0.91]        &  0.81            \\  
			$z_{\rm reio}$                 & [7.00,~15.00]       &  14.19           & [7.12,~10.25]       &  9.07            \\  
			$n_{\rm eff}$                  & [-2.40,~-2.22]      &  -2.30           & [-2.41,~-2.22]      &  -2.33           \\ 
			$f_{\rm UV}$                   & [0.00,~1.00]              &  0.02            & [0.00,~1.00]              &  0.03            \\  \hline
			$\log(f_{\rm PBH}M_{\rm{PBH}})$ & < 2.24              &  1.96            & < 1.78              &  0.34            \\ \hline\hline
			$\chi^2$/d.o.f.                &                     &  32/42           &                     &  33/43             \\ \hline
		\end{tabular}}
		\caption{$2\sigma$ limits and best fit values for the parameters of our analyses, for the two different prior choices on $z_{\rm reio}$ adopted. Values for $M_{\rm PBH}$ are expressed in units of $M_{\odot}$.\label{tab:bestfit}}
	\end{table}
	\begin{figure}
		\centering
		\includegraphics[width=0.95\linewidth]{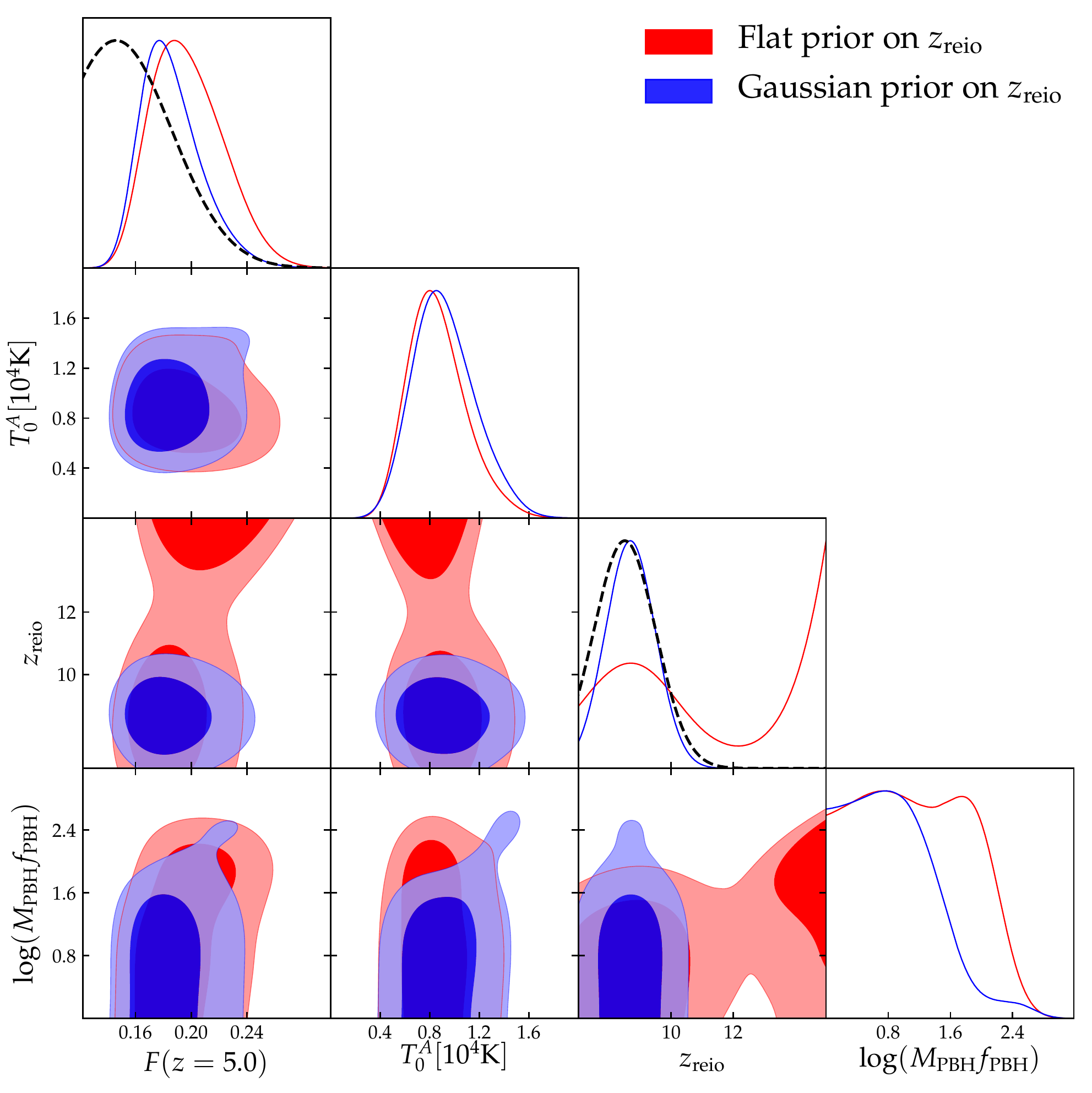}
		\caption{1 and 2$\sigma$ contour plots for some of the parameters of our analyses, for the two different prior choices on $z_{\rm reio}$. The values for $M_{\rm PBH}$ are expressed in units of $M_{\odot}$. {The dashed lines correspond to the Gaussian priors that we used for some of our analyses.}}
		\label{fig:triangle}
	\end{figure}
	
{\it Results and Discussion}.
We obtain our results by maximising a Gaussian likelihood with a Monte Carlo Markov Chain (MCMC) approach, using the publicly available MCMC sampler \texttt{emcee}~\cite{emcee13}. We adopted Gaussian priors on the mean fluxes $\bar{F}(z)$, centered on their reference values, with standard deviation $\sigma=0.04$~\cite{Irsic:2017ixq}, and on $\sigma_8$ and $n_{\rm eff}$, centered on their Planck values~\cite{Ade:2015xua}, with $\sigma = 0.05$, since the latter two parameters, whereas well constrained by CMB data, are poorly constrained by Lyman-$\alpha$ data alone~\cite{Murgia:2018now}. We adopt logarithmic priors on $f_{\rm PBH} M_{\rm PBH}$ (but our results are not affected by this choice).
%By re-running our analyses assuming a flat prior on it, the constraints are indeed unmodified, provided that a conservative Gaussian prior on the IGM temperature is also imposed, namely $T_0^A=7500 \pm 1500~{\rm{K}} \: (1 \sigma)$ (motivated by the most up-to-date IGM studies by \cite{Boera:2018vzq}). 
{Concerning the IGM thermal history, we adopt flat priors on both $T_0^A$ and $T_0^S$, in the ranges $[0,2]\cdot10^4$~K and $[-5,5]$, respectively. When the corresponding $T_0(z)$ are determined, they can assume values not enclosed by our template of simulations. When this occurs, the corresponding values of the flux power spectra are linearly extrapolated. Regarding $\gamma^S$ and $\gamma^A$, we impose flat priors on the corresponding $\gamma(z)$ (in the interval $[1,1.7]$). The priors on $z_{\rm reio}$ and $f_{\rm UV}$ are flat within the boundaries defined by our grid of simulations.}

Let us firstly focus on the simple case of PBHs featuring a MMD.
In Table~\ref{tab:bestfit} we report the marginalized $2\sigma$ constraints, in the case of a MMD, and the best fit values for all the parameters considered in our analyses. The first two columns refer to the case in which a flat prior is applied to the reionization redshift. 

The limit on the PBH abundance under the MMD assumption corresponds to:
\begin{equation}
f_{\rm PBH} M_{\rm PBH} \lesssim 170~M_{\odot}~(2\sigma),
\end{equation}
However, both Planck and~\cite{Boera:2018vzq} favour $z_{\rm reio}\sim8.5$, so we repeated our analysis with a Gaussian prior centered around $z_{\rm reio}=8.5$, with $\sigma=1.0$.
The results obtained under such assumption are shown in the last two columns of Table~\ref{tab:bestfit}, and in this case we have:
\begin{equation}
f_{\rm PBH} M_{\rm PBH} \lesssim 60~M_{\odot}~(2\sigma) \, .
\end{equation}
%This choice has been sufficient to strengthen the $2\sigma$ upper limit on $\log(M_{\rm{PBH}}/f_{\rm{PBH}})$ from 2.24 to 1.78.

\label{sec:results}

Where all DM is made by PBHs ($f_{\rm PBH} = 1$), these constraints can be interpreted as absolute limits on the PBH mass. On the other hand, such bounds weaken linearly for smaller PBH abundances ($0.05 < f_{\rm PBH} < 1$). The lower limit on $f_{\rm PBH}$ is given by the fact that, for the monochromatic case, at $z=199$,~i.e.~the redshift of the initial conditions of our simulations, if $f_{\rm PBH}$ is smaller, the Poisson effect is subdominant with respect to the so-called {\textit{seed effect}}, the treatment of which goes beyond our purposes.
\begin{figure}
	\centering
	\includegraphics[width=0.95\linewidth]{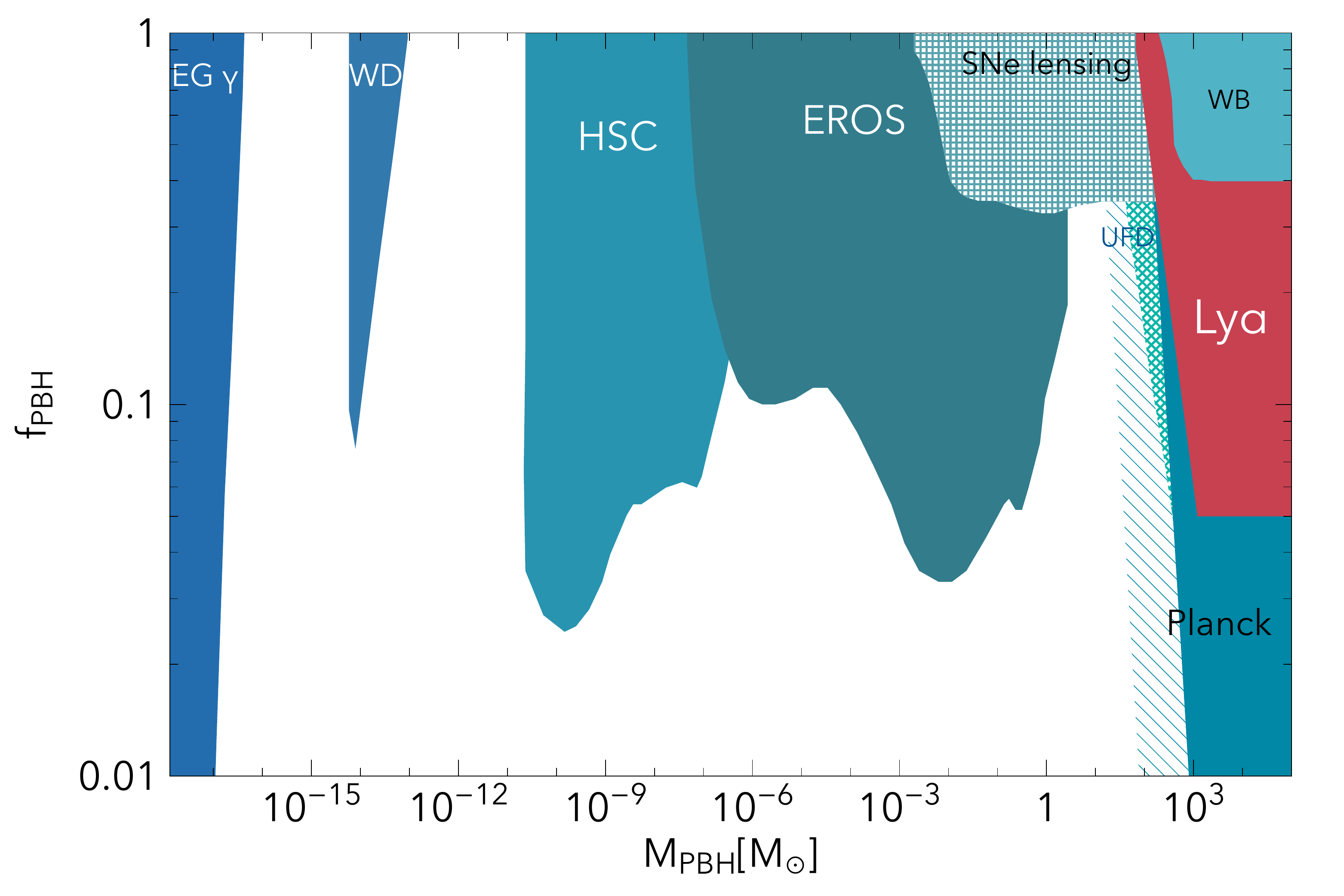}
	\includegraphics[width=1.\linewidth]{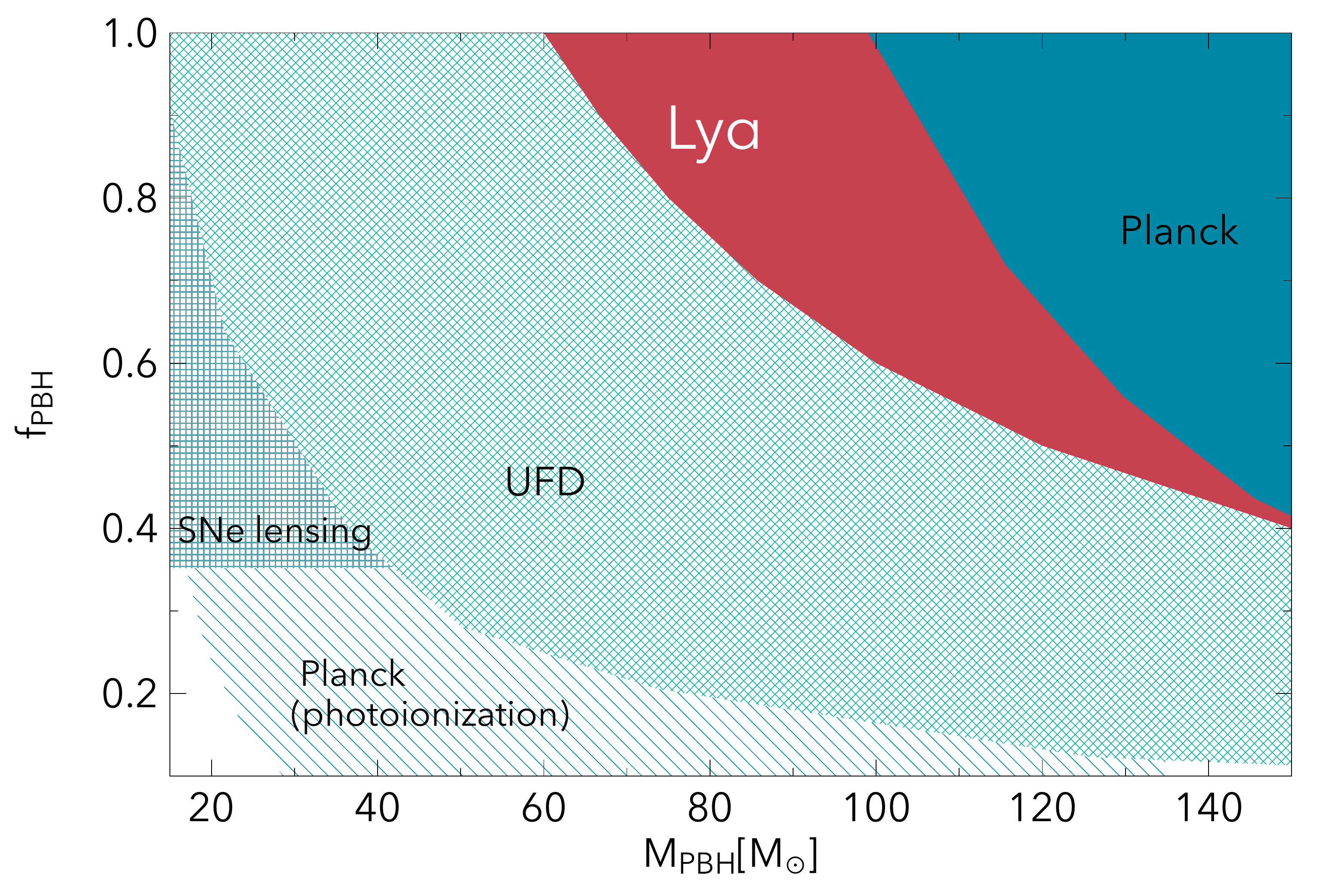}
	\caption{Present experimental constraints on the PBH abundance for MMDs (from~\cite{Bartolo:2018rku, zumalacarregui:supernovaconstraint, alihamoud:pbhaccretion,Carr:2009jm}), in shades of blue, while in red are limits from this work. Patterned areas show limits that are the most dependent on astrophysical assumptions~\cite{Carr:2016}. 
	The bottom panel is zoomed in the LIGO mass range.
	}
	\label{fig:fPBH}
\end{figure}

The degeneracy between $z_{\rm reio}$ and the PBH mass can be understood as follows: a higher reionization redshift corresponds to a more effective filtering scale, and thus to a power suppression compensated by larger values of the PBH mass.
The fact that the degeneracies are much more prominent for this parameter is telling us that the increase of power at small scales is a distinctive feature whose effect is more likely to be degenerate with a different gas filtering scale.

In Figure~\ref{fig:triangle} we show the 1 and 2$\sigma$ contours for some of the parameters of our analyses, for both prior choices on $z_{\rm reio}$.
The degeneracy between the amplitude of the IGM temperature $T_0^A(z=z_p)$ and the PBH mass derives from the opposite effects on the flux power spectra due to the increase of the two parameters. A hotter IGM implies a small-scale power suppression which is balanced by increasing $M_{\rm PBH}f_{\rm PBH}$. Slightly larger values for the mean fluxes $\bar{F}(z)$ are also required for accommodating the power enhancement induced by relatively large values of the PBH mass.
{The dashed lines represent the Gaussian priors imposed on $\bar{F}(z=5)$ and $z_{\rm reio}$, with the latter referring to the blue plots. Note that our MCMC analyses favour higher values for $\bar{F}(z=5)$ (still in agreement with its prior distribution), allowing in turn a larger power enhancement due to PBHs.} 
This is a further hint of the conservativity of the constraints presented in this work.

In order to test the stability of our results, we also performed an analysis with flat priors both on $\sigma_8$ and $n_{\rm eff}$. Under these assumptions, the constraint (using a Gaussian prior for $z_{\rm reio}$) on $f_{\rm PBH}M_{\rm PBH}$ is mildly weakened, up to 100~$M_\odot$. However, the largest values for the PBH mass are allowed only in combination with extremely low values for $n_{\rm eff}$, allowed in turn by our data set due to its poor constraining power on such parameter. As we have already stated, this is the main reason to impose a (still conservative) Gaussian prior motivated by CMB measurements on $n_{\rm eff}$.

In Figure~\ref{fig:fPBH} we report the updated plot with the constraints on the DM fraction in PBHs, in the monochromatic case. The ``LIGO window'' between $\sim20-80 M_\odot$ initially suggested in~\cite{bird:pbhasdarkmatter} has been probed and tentatively closed by constraints from Ultra-Faint Dwarf galaxies~\cite{brandt:ufdgconstraint} and Supernov\ae~lensing~\cite{zumalacarregui:supernovaconstraint}; these constraints have been questioned because of astrophysics uncertainties (e.g.,~\cite{Li:2016, Carr:2016}~{\footnote{See also~{\href{https://indico.cern.ch/event/686745/}{Primordial versus Astrophysical Origin of Black Holes -- CERN workshop}}}}): we show them in a patterned area. In this work we robustly close the higher mass part of that remaining window. There remains however, an interesting possibility in the very low mass range, $\lesssim10^{-10}M_\odot$~\cite{Pi:2017, Bartolo:2018rku}.

\begin{figure}[t]
	\includegraphics[width=0.95\linewidth]{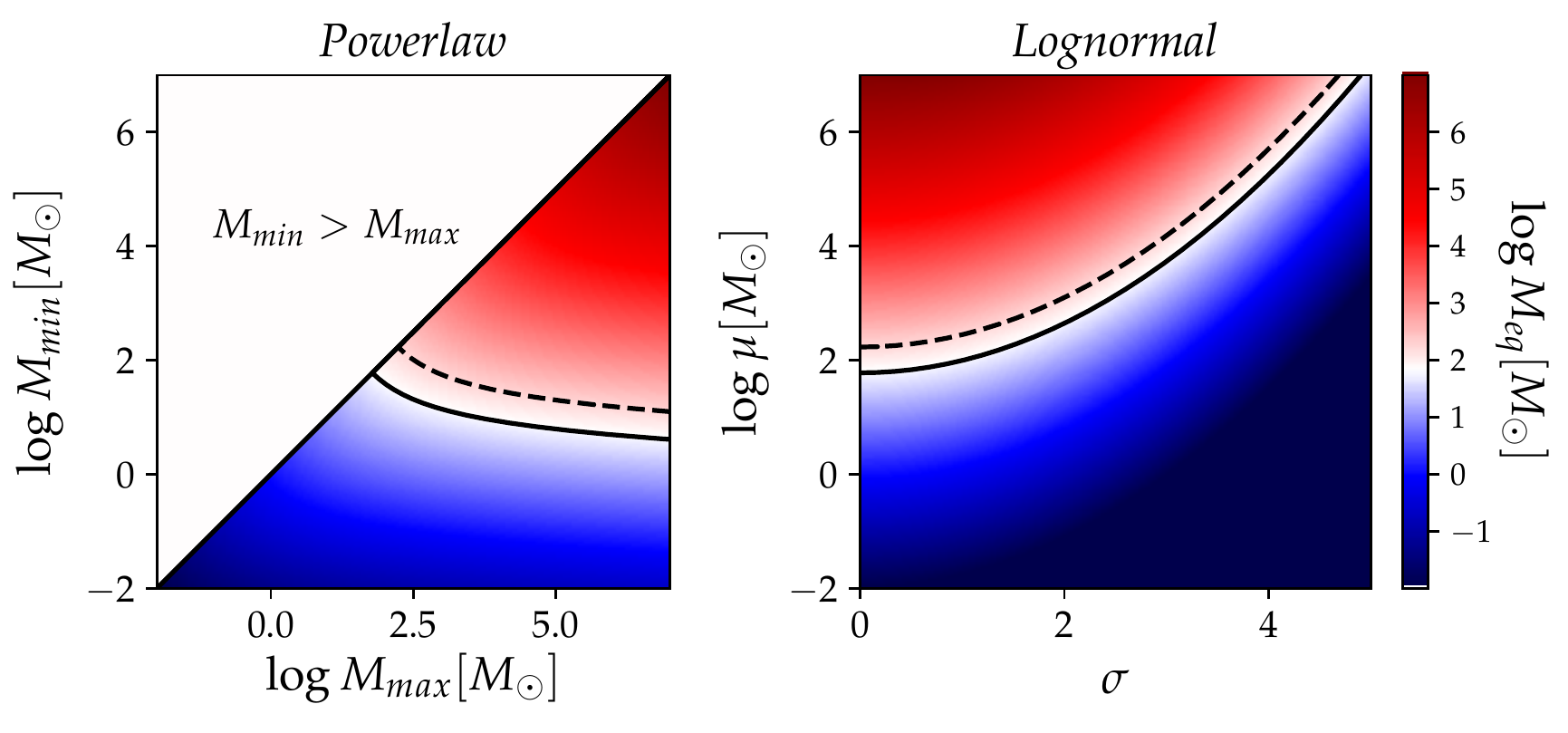}
	\caption{Equivalent Mass $M_{\mathrm{eq}}$ for EMDs.
		\textit{Left: Powerlaw} with $M_{\mathrm{min}}, \: M_{\mathrm{max}} \in [10^{-2},10^{7}] \: M_{\odot}$, and $\tilde\gamma = 0$. \textit{Right: Lognormal} with $\mu \in [10^{-2},10^{7}] \: M_{\odot}$, $\sigma \in [0,5]$. Solid lines are $M_{\mathrm{eq}} = 60  \: M_{\odot}$, dashed lines refer to $M_{\mathrm{eq}} = 170$.} 
		%Red regions are thereby ruled out by our analyses, whereas  blue ones are admitted.} 
	\label{fig:EMD_constraints}
\end{figure}

By defining an equivalent mass $M_{\rm eq}$ one can convert the limits for the MMD case to bounds on the parameters of a given EMD.
In Figure~\ref{fig:EMD_constraints} we provide such bounds, {similarly to what was shown in Figure~3~of~\cite{Carr:2017jsz} for other observational constraints}.
In other words, each of the panels maps the limits on MMDs peaked at $M_{\mathrm{eq}}$ to constraints on EMDs.
The left panel shows the~\textit{Powerlaw} EMD, with $\tilde{\gamma} = 0$, focusing on the following mass range: $M_{\mathrm{min}}, \: M_{\mathrm{max}} \in [10^{-2},10^{7}] \: M_{\odot}$.
In the right panel we show the~\textit{Lognormal} EMD, scanning the parameter space defined by $\mu \in [10^{-2},10^{7}] \: M_{\odot}$, and $\sigma \in [0,5]$. 
The two black lines correspond to the contraints quoted above,~i.e.~$M_{\mathrm{eq}} = 60  \: M_{\odot}$~($solid$), and $M_{\mathrm{eq}} = 170  \: M_{\odot}$~($dashed$). The blue regions are admitted by our analyses, while the red areas ruled out.
All our analyses are based on the straightforward assumption that the PBH number density $n_{\mathrm{PBH}}$ is fixed during the cosmic time investigated by our simulations,~i.e.~from $z=199$ to $z=4.2$, which indeed corresponds to epochs when PBHs do not form anymore.
However, the possibility that PBHs can fully evaporate during such time interval would alter our conclusions, since in that case $n_{\rm {PBH}}$ would vary with time. Nevertheless, from $z=199$ to now, only PBHs with masses smaller than $\mathcal{O}(10^{-18})~M_{\odot}$ might completely evaporate~(\cite{hawking:1974radiation}). For this reason, such value has to lie below the PBH mass ranges investigated in this work. We have thus set $10^{-18}~M_{\odot}$ as lower limit for the integral in Equation~\eqref{eq:eq_mass}, and $10^{7}~M_{\odot}$ as upper limit. \\ \vspace{-0.25cm}

{\it Conclusions}.
\label{sec:concl}
In this work we have presented new bounds on the DM fraction in PBHs, using an extensive analysis of high-redshift Lyman-$\alpha$ forest data, improving over previous similar analyses in three different ways: 1) we used the high-resolution MIKE/HIRES data, exploring better the high-redshift range where primordial differences are more prominent; 2) we relied on very accurate high-resolution hydrodynamic simulations which expands over a thermal history suggested by data; 3) we used the full shape of the 1D flux power rather than a single amplitude parameter.   

% We have stressed the dependence on such limits on the reionization history of the universe.
% However, under both the considered assumptions on the reionization redshift, we have improved the previously published constraints, extracted from the same astrophysical observable, nearly by 3 orders of magnitude. 
Our results improve previous constraints by roughly 2 orders of magnitude; furthermore, we have generalized our results to non-monochromatic PBH mass distributions, and ruled out a large part of the parameter space for two of the most popular EMDs: {\textit{Powerlaw}} and {\textit{Lognormal}}.

In the near future, it is expected that a larger number of high-redshift, high-resolution and signal-to-noise quasar spectra collected with the ESPRESSO spectrograph~\cite{pepe13} or at the E-ELT could allow to achieve tighter constraints.
Another relevant aspect would be an accurate modeling of the heating and ionization due to accretion effects around the PBHs, to quantify how and if they could impact on the (much larger) scales of the Lyman-$\alpha$ forest.

{Whereas PBHs with mass $\mathcal{O}(10)$ can potentially solve some tensions in the cosmic infrared background~\cite{Kashlinsky:2016sdv,Kashlinsky:2004jt,Kashlinsky:2005di},}
the accumulation of limits on the PBHs as DM model in the mass range probed by LIGO seems to suggest that the hypothesis of $30 M_\odot$ PBHs being the DM is less and less likely to be true.

It has however become clear that these studies brought a plethora of astrophysical information, and even the exclusion of certain PBH mass ranges will bring information on some of the processes happening in the very early Universe. \\

%\appendix
%\section{Some title}

{\it Acknowledgments}.
The authors are thankful to Nils Sch$\ddot{\rm o}$neberg, Nicola Bellomo, Jos\'e Luis Bernal, Licia Verde, Pasquale D. Serpico, Takeshi Kobayashi, Vid Ir\v{s}i\v{c}, Emiliano Sefusatti, and Gabrijela Zaharijas for helpful discussions. RM, GS, and MV are supported by the INFN INDARK PD51 grant. The simulations were performed on the Ulysses SISSA/ICTP supercomputer.

%%%%%%%%%%%%%%%%%%%%%%%%%%%%%%%%%%%%%%%%%%%%%%%%%%



%\bibliographystyle{mnras}
\bibliography{wdm2-1} 

\begin{thebibliography}{88}%
\makeatletter
\providecommand \@ifxundefined [1]{%
 \@ifx{#1\undefined}
}%
\providecommand \@ifnum [1]{%
 \ifnum #1\expandafter \@firstoftwo
 \else \expandafter \@secondoftwo
 \fi
}%
\providecommand \@ifx [1]{%
 \ifx #1\expandafter \@firstoftwo
 \else \expandafter \@secondoftwo
 \fi
}%
\providecommand \natexlab [1]{#1}%
\providecommand \enquote  [1]{``#1''}%
\providecommand \bibnamefont  [1]{#1}%
\providecommand \bibfnamefont [1]{#1}%
\providecommand \citenamefont [1]{#1}%
\providecommand \href@noop [0]{\@secondoftwo}%
\providecommand \href [0]{\begingroup \@sanitize@url \@href}%
\providecommand \@href[1]{\@@startlink{#1}\@@href}%
\providecommand \@@href[1]{\endgroup#1\@@endlink}%
\providecommand \@sanitize@url [0]{\catcode `\\12\catcode `\$12\catcode
  `\&12\catcode `\#12\catcode `\^12\catcode `\_12\catcode `\%12\relax}%
\providecommand \@@startlink[1]{}%
\providecommand \@@endlink[0]{}%
\providecommand \url  [0]{\begingroup\@sanitize@url \@url }%
\providecommand \@url [1]{\endgroup\@href {#1}{\urlprefix }}%
\providecommand \urlprefix  [0]{URL }%
\providecommand \Eprint [0]{\href }%
\providecommand \doibase [0]{http://dx.doi.org/}%
\providecommand \selectlanguage [0]{\@gobble}%
\providecommand \bibinfo  [0]{\@secondoftwo}%
\providecommand \bibfield  [0]{\@secondoftwo}%
\providecommand \translation [1]{[#1]}%
\providecommand \BibitemOpen [0]{}%
\providecommand \bibitemStop [0]{}%
\providecommand \bibitemNoStop [0]{.\EOS\space}%
\providecommand \EOS [0]{\spacefactor3000\relax}%
\providecommand \BibitemShut  [1]{\csname bibitem#1\endcsname}%
\let\auto@bib@innerbib\@empty
%</preamble>
\bibitem [{\citenamefont {Hawking}(1971)}]{Hawking:pbh1971}%
  \BibitemOpen
  \bibfield  {author} {\bibinfo {author} {\bibfnamefont {S.}~\bibnamefont
  {Hawking}},\ }\href {\doibase 10.1093/mnras/152.1.75} {\bibfield  {journal}
  {\bibinfo  {journal} {MNRAS}\ }\textbf {\bibinfo {volume} {152}},\ \bibinfo
  {pages} {75} (\bibinfo {year} {1971})}\BibitemShut {NoStop}%
\bibitem [{\citenamefont {Ivanov}\ \emph {et~al.}(1994)\citenamefont {Ivanov},
  \citenamefont {Naselsky},\ and\ \citenamefont
  {Novikov}}]{ivanov:pbhfrominflationI}%
  \BibitemOpen
  \bibfield  {author} {\bibinfo {author} {\bibfnamefont {P.}~\bibnamefont
  {Ivanov}}, \bibinfo {author} {\bibfnamefont {P.}~\bibnamefont {Naselsky}}, \
  and\ \bibinfo {author} {\bibfnamefont {I.}~\bibnamefont {Novikov}},\ }\href
  {\doibase 10.1103/PhysRevD.50.7173} {\bibfield  {journal} {\bibinfo
  {journal} {Phys. Rev. D}\ }\textbf {\bibinfo {volume} {50}},\ \bibinfo
  {pages} {7173} (\bibinfo {year} {1994})}\BibitemShut {NoStop}%
\bibitem [{\citenamefont {Garc\'{\i}a-Bellido}\ \emph
  {et~al.}(1996)\citenamefont {Garc\'{\i}a-Bellido}, \citenamefont {Linde},\
  and\ \citenamefont {Wands}}]{Bellido:pbh}%
  \BibitemOpen
  \bibfield  {author} {\bibinfo {author} {\bibfnamefont {J.}~\bibnamefont
  {Garc\'{\i}a-Bellido}}, \bibinfo {author} {\bibfnamefont {A.}~\bibnamefont
  {Linde}}, \ and\ \bibinfo {author} {\bibfnamefont {D.}~\bibnamefont
  {Wands}},\ }\href {\doibase 10.1103/PhysRevD.54.6040} {\bibfield  {journal}
  {\bibinfo  {journal} {Phys. Rev. D}\ }\textbf {\bibinfo {volume} {54}},\
  \bibinfo {pages} {6040} (\bibinfo {year} {1996})},\ \Eprint
  {http://arxiv.org/abs/astro-ph/9605094} {arXiv:astro-ph/9605094} \BibitemShut
  {NoStop}%
\bibitem [{\citenamefont {Ivanov}(1998)}]{ivanov:pbhfrominflationII}%
  \BibitemOpen
  \bibfield  {author} {\bibinfo {author} {\bibfnamefont {P.}~\bibnamefont
  {Ivanov}},\ }\href {\doibase 10.1103/PhysRevD.57.7145} {\bibfield  {journal}
  {\bibinfo  {journal} {Phys. Rev. D}\ }\textbf {\bibinfo {volume} {57}},\
  \bibinfo {pages} {7145} (\bibinfo {year} {1998})},\ \Eprint
  {http://arxiv.org/abs/astro-ph/9708224} {arXiv:astro-ph/9708224} \BibitemShut
  {NoStop}%
\bibitem [{\citenamefont {Polnarev}\ and\ \citenamefont
  {Zembowicz}(1991)}]{Polnarev88}%
  \BibitemOpen
  \bibfield  {author} {\bibinfo {author} {\bibfnamefont {A.}~\bibnamefont
  {Polnarev}}\ and\ \bibinfo {author} {\bibfnamefont {R.}~\bibnamefont
  {Zembowicz}},\ }\href {\doibase 10.1103/PhysRevD.43.1106} {\bibfield
  {journal} {\bibinfo  {journal} {Phys. Rev. D}\ }\textbf {\bibinfo {volume}
  {43}},\ \bibinfo {pages} {1106} (\bibinfo {year} {1991})}\BibitemShut
  {NoStop}%
\bibitem [{\citenamefont {Hawking}(1989)}]{HAWKING1989237}%
  \BibitemOpen
  \bibfield  {author} {\bibinfo {author} {\bibfnamefont {S.}~\bibnamefont
  {Hawking}},\ }\href {\doibase https://doi.org/10.1016/0370-2693(89)90206-2}
  {\bibfield  {journal} {\bibinfo  {journal} {Physics Letters B}\ }\textbf
  {\bibinfo {volume} {231}},\ \bibinfo {pages} {237 } (\bibinfo {year}
  {1989})}\BibitemShut {NoStop}%
\bibitem [{\citenamefont {Wichoski}\ \emph {et~al.}(1998)\citenamefont
  {Wichoski}, \citenamefont {MacGibbon},\ and\ \citenamefont
  {Brandenberger}}]{WICHOSKI1998191}%
  \BibitemOpen
  \bibfield  {author} {\bibinfo {author} {\bibfnamefont {U.~F.}\ \bibnamefont
  {Wichoski}}, \bibinfo {author} {\bibfnamefont {J.~H.}\ \bibnamefont
  {MacGibbon}}, \ and\ \bibinfo {author} {\bibfnamefont {R.~H.}\ \bibnamefont
  {Brandenberger}},\ }\href {\doibase 10.1016/S0370-1573(98)00070-2} {\bibfield
   {journal} {\bibinfo  {journal} {Physics Reports}\ }\textbf {\bibinfo
  {volume} {307}},\ \bibinfo {pages} {191 } (\bibinfo {year}
  {1998})}\BibitemShut {NoStop}%
\bibitem [{\citenamefont {Berezin}\ \emph {et~al.}(1983)\citenamefont
  {Berezin}, \citenamefont {Kuzmin},\ and\ \citenamefont
  {Tkachev}}]{BEREZIN198391}%
  \BibitemOpen
  \bibfield  {author} {\bibinfo {author} {\bibfnamefont {V.}~\bibnamefont
  {Berezin}}, \bibinfo {author} {\bibfnamefont {V.}~\bibnamefont {Kuzmin}}, \
  and\ \bibinfo {author} {\bibfnamefont {I.}~\bibnamefont {Tkachev}},\ }\href
  {\doibase https://doi.org/10.1016/0370-2693(83)90630-5} {\bibfield  {journal}
  {\bibinfo  {journal} {Physics Letters B}\ }\textbf {\bibinfo {volume}
  {120}},\ \bibinfo {pages} {91 } (\bibinfo {year} {1983})}\BibitemShut
  {NoStop}%
\bibitem [{\citenamefont {Ipser}\ and\ \citenamefont
  {Sikivie}(1984)}]{Ipser84}%
  \BibitemOpen
  \bibfield  {author} {\bibinfo {author} {\bibfnamefont {J.}~\bibnamefont
  {Ipser}}\ and\ \bibinfo {author} {\bibfnamefont {P.}~\bibnamefont
  {Sikivie}},\ }\href {\doibase 10.1103/PhysRevD.30.712} {\bibfield  {journal}
  {\bibinfo  {journal} {Phys. Rev. D}\ }\textbf {\bibinfo {volume} {30}},\
  \bibinfo {pages} {712} (\bibinfo {year} {1984})}\BibitemShut {NoStop}%
\bibitem [{\citenamefont {Crawford}\ and\ \citenamefont
  {Schramm}(1982)}]{Crawford82}%
  \BibitemOpen
  \bibfield  {author} {\bibinfo {author} {\bibfnamefont {M.}~\bibnamefont
  {Crawford}}\ and\ \bibinfo {author} {\bibfnamefont {D.~N.}\ \bibnamefont
  {Schramm}},\ }\href {\doibase http://dx.doi.org/10.1038/298538a0} {\bibfield
  {journal} {\bibinfo  {journal} {Nature}\ }\textbf {\bibinfo {volume} {298}},\
  \bibinfo {pages} {538} (\bibinfo {year} {1982})}\BibitemShut {NoStop}%
\bibitem [{\citenamefont {La}\ and\ \citenamefont
  {Steinhardt}(1989)}]{LA1989375}%
  \BibitemOpen
  \bibfield  {author} {\bibinfo {author} {\bibfnamefont {D.}~\bibnamefont
  {La}}\ and\ \bibinfo {author} {\bibfnamefont {P.~J.}\ \bibnamefont
  {Steinhardt}},\ }\href {\doibase
  https://doi.org/10.1016/0370-2693(89)90890-3} {\bibfield  {journal} {\bibinfo
   {journal} {Physics Letters B}\ }\textbf {\bibinfo {volume} {220}},\ \bibinfo
  {pages} {375 } (\bibinfo {year} {1989})}\BibitemShut {NoStop}%
\bibitem [{\citenamefont {Shandera}\ \emph {et~al.}(2018)\citenamefont
  {Shandera}, \citenamefont {Jeong},\ and\ \citenamefont
  {Gebhardt}}]{shandera:pbhfromdarkmatter}%
  \BibitemOpen
  \bibfield  {author} {\bibinfo {author} {\bibfnamefont {S.}~\bibnamefont
  {Shandera}}, \bibinfo {author} {\bibfnamefont {D.}~\bibnamefont {Jeong}}, \
  and\ \bibinfo {author} {\bibfnamefont {H.~S.~G.}\ \bibnamefont {Gebhardt}},\
  }\href {\doibase 10.1103/PhysRevLett.120.241102} {\bibfield  {journal}
  {\bibinfo  {journal} {Phys. Rev. Lett.}\ }\textbf {\bibinfo {volume} {120}},\
  \bibinfo {pages} {241102} (\bibinfo {year} {2018})},\ \Eprint
  {http://arxiv.org/abs/1802.08206} {arXiv:1802.08206} \BibitemShut {NoStop}%
\bibitem [{\citenamefont {Abbott}\ \emph
  {et~al.}(2016{\natexlab{a}})\citenamefont {Abbott} \emph
  {et~al.}}]{abbott:firstligodetection}%
  \BibitemOpen
  \bibfield  {author} {\bibinfo {author} {\bibfnamefont {B.~P.}\ \bibnamefont
  {Abbott}} \emph {et~al.} (\bibinfo {collaboration} {LIGO Scientific
  Collaboration and Virgo Collaboration}),\ }\href {\doibase
  10.1103/PhysRevLett.116.061102} {\bibfield  {journal} {\bibinfo  {journal}
  {Phys. Rev. Lett.}\ }\textbf {\bibinfo {volume} {116}},\ \bibinfo {pages}
  {061102} (\bibinfo {year} {2016}{\natexlab{a}})},\ \Eprint
  {http://arxiv.org/abs/1602.03837} {arXiv:1602.03837} \BibitemShut {NoStop}%
\bibitem [{\citenamefont {Abbott}\ \emph
  {et~al.}(2016{\natexlab{b}})\citenamefont {Abbott} \emph
  {et~al.}}]{abbott:firstligodetectionproperties}%
  \BibitemOpen
  \bibfield  {author} {\bibinfo {author} {\bibfnamefont {B.~P.}\ \bibnamefont
  {Abbott}} \emph {et~al.} (\bibinfo {collaboration} {LIGO Scientific
  Collaboration and Virgo Collaboration}),\ }\href {\doibase
  10.1103/PhysRevLett.116.241102} {\bibfield  {journal} {\bibinfo  {journal}
  {Phys. Rev. Lett.}\ }\textbf {\bibinfo {volume} {116}},\ \bibinfo {pages}
  {241102} (\bibinfo {year} {2016}{\natexlab{b}})},\ \Eprint
  {http://arxiv.org/abs/1602.03840} {arXiv:1602.03840} \BibitemShut {NoStop}%
\bibitem [{\citenamefont {Bird}\ \emph {et~al.}(2016)\citenamefont {Bird},
  \citenamefont {Cholis}, \citenamefont {Mu\~noz}, \citenamefont
  {Ali-Ha\"{\i}moud}, \citenamefont {Kamionkowski}, \citenamefont {Kovetz},
  \citenamefont {Raccanelli},\ and\ \citenamefont
  {Riess}}]{bird:pbhasdarkmatter}%
  \BibitemOpen
  \bibfield  {author} {\bibinfo {author} {\bibfnamefont {S.}~\bibnamefont
  {Bird}}, \bibinfo {author} {\bibfnamefont {I.}~\bibnamefont {Cholis}},
  \bibinfo {author} {\bibfnamefont {J.~B.}\ \bibnamefont {Mu\~noz}}, \bibinfo
  {author} {\bibfnamefont {Y.}~\bibnamefont {Ali-Ha\"{\i}moud}}, \bibinfo
  {author} {\bibfnamefont {M.}~\bibnamefont {Kamionkowski}}, \bibinfo {author}
  {\bibfnamefont {E.~D.}\ \bibnamefont {Kovetz}}, \bibinfo {author}
  {\bibfnamefont {A.}~\bibnamefont {Raccanelli}}, \ and\ \bibinfo {author}
  {\bibfnamefont {A.~G.}\ \bibnamefont {Riess}},\ }\href {\doibase
  10.1103/PhysRevLett.116.201301} {\bibfield  {journal} {\bibinfo  {journal}
  {Phys. Rev. Lett.}\ }\textbf {\bibinfo {volume} {116}},\ \bibinfo {pages}
  {201301} (\bibinfo {year} {2016})},\ \Eprint
  {http://arxiv.org/abs/1603.00464} {arXiv:1603.00464} \BibitemShut {NoStop}%
\bibitem [{\citenamefont {Raccanelli}\ \emph {et~al.}(2016)\citenamefont
  {Raccanelli}, \citenamefont {Kovetz}, \citenamefont {Bird}, \citenamefont
  {Cholis},\ and\ \citenamefont {Mu\~noz}}]{raccanelli:pbhprogenitors}%
  \BibitemOpen
  \bibfield  {author} {\bibinfo {author} {\bibfnamefont {A.}~\bibnamefont
  {Raccanelli}}, \bibinfo {author} {\bibfnamefont {E.~D.}\ \bibnamefont
  {Kovetz}}, \bibinfo {author} {\bibfnamefont {S.}~\bibnamefont {Bird}},
  \bibinfo {author} {\bibfnamefont {I.}~\bibnamefont {Cholis}}, \ and\ \bibinfo
  {author} {\bibfnamefont {J.~B.}\ \bibnamefont {Mu\~noz}},\ }\href {\doibase
  10.1103/PhysRevD.94.023516} {\bibfield  {journal} {\bibinfo  {journal} {Phys.
  Rev. D}\ }\textbf {\bibinfo {volume} {94}},\ \bibinfo {pages} {023516}
  (\bibinfo {year} {2016})},\ \Eprint {http://arxiv.org/abs/1605.01405}
  {arXiv:1605.01405} \BibitemShut {NoStop}%
\bibitem [{\citenamefont {Scelfo}\ \emph {et~al.}(2018)\citenamefont {Scelfo},
  \citenamefont {Bellomo}, \citenamefont {Raccanelli}, \citenamefont
  {Matarrese},\ and\ \citenamefont {Verde}}]{Scelfo_2018}%
  \BibitemOpen
  \bibfield  {author} {\bibinfo {author} {\bibfnamefont {G.}~\bibnamefont
  {Scelfo}}, \bibinfo {author} {\bibfnamefont {N.}~\bibnamefont {Bellomo}},
  \bibinfo {author} {\bibfnamefont {A.}~\bibnamefont {Raccanelli}}, \bibinfo
  {author} {\bibfnamefont {S.}~\bibnamefont {Matarrese}}, \ and\ \bibinfo
  {author} {\bibfnamefont {L.}~\bibnamefont {Verde}},\ }\href {\doibase
  10.1088/1475-7516/2018/09/039} {\bibfield  {journal} {\bibinfo  {journal}
  {JCAP}\ }\textbf {\bibinfo {volume} {2018}},\ \bibinfo {pages} {039}
  (\bibinfo {year} {2018})},\ \Eprint {http://arxiv.org/abs/1809.03528v1}
  {arXiv:1809.03528v1} \BibitemShut {NoStop}%
\bibitem [{\citenamefont {Cholis}\ \emph {et~al.}(2016)\citenamefont {Cholis},
  \citenamefont {Kovetz}, \citenamefont {Ali-Ha{\"\i}moud}, \citenamefont
  {Bird}, \citenamefont {Kamionkowski}, \citenamefont {Mu{\~n}oz},\ and\
  \citenamefont {Raccanelli}}]{cholis:orbitaleccentricities}%
  \BibitemOpen
  \bibfield  {author} {\bibinfo {author} {\bibfnamefont {I.}~\bibnamefont
  {Cholis}}, \bibinfo {author} {\bibfnamefont {E.~D.}\ \bibnamefont {Kovetz}},
  \bibinfo {author} {\bibfnamefont {Y.}~\bibnamefont {Ali-Ha{\"\i}moud}},
  \bibinfo {author} {\bibfnamefont {S.}~\bibnamefont {Bird}}, \bibinfo {author}
  {\bibfnamefont {M.}~\bibnamefont {Kamionkowski}}, \bibinfo {author}
  {\bibfnamefont {J.~B.}\ \bibnamefont {Mu{\~n}oz}}, \ and\ \bibinfo {author}
  {\bibfnamefont {A.}~\bibnamefont {Raccanelli}},\ }\href {\doibase
  10.1103/PhysRevD.94.084013} {\bibfield  {journal} {\bibinfo  {journal} {Phys.
  Rev.}\ }\textbf {\bibinfo {volume} {D94}},\ \bibinfo {pages} {084013}
  (\bibinfo {year} {2016})},\ \Eprint {http://arxiv.org/abs/1606.07437}
  {arXiv:1606.07437 [astro-ph.HE]} \BibitemShut {NoStop}%
%%CITATION = ARXIV:1606.07437;%%
\bibitem [{\citenamefont {Kovetz}\ \emph {et~al.}(2017)\citenamefont {Kovetz},
  \citenamefont {Cholis}, \citenamefont {Breysse},\ and\ \citenamefont
  {Kamionkowski}}]{kovetz:pbhmassfunction}%
  \BibitemOpen
  \bibfield  {author} {\bibinfo {author} {\bibfnamefont {E.~D.}\ \bibnamefont
  {Kovetz}}, \bibinfo {author} {\bibfnamefont {I.}~\bibnamefont {Cholis}},
  \bibinfo {author} {\bibfnamefont {P.~C.}\ \bibnamefont {Breysse}}, \ and\
  \bibinfo {author} {\bibfnamefont {M.}~\bibnamefont {Kamionkowski}},\ }\href
  {\doibase 10.1103/PhysRevD.95.103010} {\bibfield  {journal} {\bibinfo
  {journal} {Phys. Rev.}\ }\textbf {\bibinfo {volume} {D95}},\ \bibinfo {pages}
  {103010} (\bibinfo {year} {2017})},\ \Eprint
  {http://arxiv.org/abs/1611.01157} {arXiv:1611.01157 [astro-ph.CO]}
  \BibitemShut {NoStop}%
%%CITATION = ARXIV:1611.01157;%%
\bibitem [{\citenamefont {Kovetz}(2017)}]{kovetz:pbhandgw}%
  \BibitemOpen
  \bibfield  {author} {\bibinfo {author} {\bibfnamefont {E.~D.}\ \bibnamefont
  {Kovetz}},\ }\href {\doibase 10.1103/PhysRevLett.119.131301} {\bibfield
  {journal} {\bibinfo  {journal} {Phys. Rev. Lett.}\ }\textbf {\bibinfo
  {volume} {119}},\ \bibinfo {pages} {131301} (\bibinfo {year} {2017})},\
  \Eprint {http://arxiv.org/abs/1705.09182} {arXiv:1705.09182} \BibitemShut
  {NoStop}%
\bibitem [{\citenamefont {Mu{\~n}oz}\ \emph {et~al.}(2016)\citenamefont
  {Mu{\~n}oz}, \citenamefont {Kovetz}, \citenamefont {Dai},\ and\ \citenamefont
  {Kamionkowski}}]{munoz:fastradioburst}%
  \BibitemOpen
  \bibfield  {author} {\bibinfo {author} {\bibfnamefont {J.~B.}\ \bibnamefont
  {Mu{\~n}oz}}, \bibinfo {author} {\bibfnamefont {E.~D.}\ \bibnamefont
  {Kovetz}}, \bibinfo {author} {\bibfnamefont {L.}~\bibnamefont {Dai}}, \ and\
  \bibinfo {author} {\bibfnamefont {M.}~\bibnamefont {Kamionkowski}},\ }\href
  {\doibase 10.1103/PhysRevLett.117.091301} {\bibfield  {journal} {\bibinfo
  {journal} {Phys. Rev. Lett.}\ }\textbf {\bibinfo {volume} {117}},\ \bibinfo
  {pages} {091301} (\bibinfo {year} {2016})},\ \Eprint
  {http://arxiv.org/abs/1605.00008} {arXiv:1605.00008} \BibitemShut {NoStop}%
\bibitem [{\citenamefont {Barnacka}\ \emph {et~al.}(2012)\citenamefont
  {Barnacka}, \citenamefont {Glicenstein},\ and\ \citenamefont
  {Moderski}}]{barnacka:femtolensingconstraint}%
  \BibitemOpen
  \bibfield  {author} {\bibinfo {author} {\bibfnamefont {A.}~\bibnamefont
  {Barnacka}}, \bibinfo {author} {\bibfnamefont {J.-F.}\ \bibnamefont
  {Glicenstein}}, \ and\ \bibinfo {author} {\bibfnamefont {R.}~\bibnamefont
  {Moderski}},\ }\href {\doibase 10.1103/PhysRevD.86.043001} {\bibfield
  {journal} {\bibinfo  {journal} {Phys. Rev. D}\ }\textbf {\bibinfo {volume}
  {86}},\ \bibinfo {pages} {043001} (\bibinfo {year} {2012})},\ \Eprint
  {http://arxiv.org/abs/1204.2056} {arXiv:1204.2056} \BibitemShut {NoStop}%
\bibitem [{\citenamefont {Katz}\ \emph {et~al.}(2018)\citenamefont {Katz},
  \citenamefont {Kopp}, \citenamefont {Sibiryakov},\ and\ \citenamefont
  {Xue}}]{katz:femtolensingconstraint}%
  \BibitemOpen
  \bibfield  {author} {\bibinfo {author} {\bibfnamefont {A.}~\bibnamefont
  {Katz}}, \bibinfo {author} {\bibfnamefont {J.}~\bibnamefont {Kopp}}, \bibinfo
  {author} {\bibfnamefont {S.}~\bibnamefont {Sibiryakov}}, \ and\ \bibinfo
  {author} {\bibfnamefont {W.}~\bibnamefont {Xue}},\ }\href {\doibase
  10.1088/1475-7516/2018/12/005} {\bibfield  {journal} {\bibinfo  {journal}
  {JCAP}\ }\textbf {\bibinfo {volume} {1812}},\ \bibinfo {pages} {005}
  (\bibinfo {year} {2018})},\ \Eprint {http://arxiv.org/abs/1807.11495}
  {arXiv:1807.11495 [astro-ph.CO]} \BibitemShut {NoStop}%
%%CITATION = ARXIV:1807.11495;%%
\bibitem [{\citenamefont {Griest}\ \emph {et~al.}(2014)\citenamefont {Griest},
  \citenamefont {Cieplak},\ and\ \citenamefont
  {Lehner}}]{griest:keplerconstraint}%
  \BibitemOpen
  \bibfield  {author} {\bibinfo {author} {\bibfnamefont {K.}~\bibnamefont
  {Griest}}, \bibinfo {author} {\bibfnamefont {A.~M.}\ \bibnamefont {Cieplak}},
  \ and\ \bibinfo {author} {\bibfnamefont {M.~J.}\ \bibnamefont {Lehner}},\
  }\href {\doibase 10.1088/0004-637X/786/2/158} {\bibfield  {journal} {\bibinfo
   {journal} {\apj}\ }\textbf {\bibinfo {volume} {786}},\ \bibinfo {pages}
  {158} (\bibinfo {year} {2014})},\ \Eprint {http://arxiv.org/abs/1307.5798}
  {arXiv:1307.5798} \BibitemShut {NoStop}%
\bibitem [{\citenamefont {Niikura}\ \emph {et~al.}(2017)\citenamefont
  {Niikura}, \citenamefont {Takada}, \citenamefont {Yasuda}, \citenamefont
  {Lupton}, \citenamefont {Sumi}, \citenamefont {More}, \citenamefont {More},
  \citenamefont {Oguri},\ and\ \citenamefont
  {Chiba}}]{niikura:microlensingconstraint}%
  \BibitemOpen
  \bibfield  {author} {\bibinfo {author} {\bibfnamefont {H.}~\bibnamefont
  {Niikura}}, \bibinfo {author} {\bibfnamefont {M.}~\bibnamefont {Takada}},
  \bibinfo {author} {\bibfnamefont {N.}~\bibnamefont {Yasuda}}, \bibinfo
  {author} {\bibfnamefont {R.~H.}\ \bibnamefont {Lupton}}, \bibinfo {author}
  {\bibfnamefont {T.}~\bibnamefont {Sumi}}, \bibinfo {author} {\bibfnamefont
  {S.}~\bibnamefont {More}}, \bibinfo {author} {\bibfnamefont {A.}~\bibnamefont
  {More}}, \bibinfo {author} {\bibfnamefont {M.}~\bibnamefont {Oguri}}, \ and\
  \bibinfo {author} {\bibfnamefont {M.}~\bibnamefont {Chiba}},\ }\href@noop {}
  {\bibfield  {journal} {\bibinfo  {journal} {1701.02151}\ } (\bibinfo {year}
  {2017})}\BibitemShut {NoStop}%
\bibitem [{\citenamefont {Tisserand}\ \emph {et~al.}(2007)\citenamefont
  {Tisserand} \emph {et~al.}}]{tisserand:microlensingconstraint}%
  \BibitemOpen
  \bibfield  {author} {\bibinfo {author} {\bibfnamefont {P.}~\bibnamefont
  {Tisserand}} \emph {et~al.} (\bibinfo {collaboration} {EROS-2
  Collaboration}),\ }\href {\doibase 10.1051/0004-6361:20066017} {\bibfield
  {journal} {\bibinfo  {journal} {A\&A}\ }\textbf {\bibinfo {volume} {469}},\
  \bibinfo {pages} {387} (\bibinfo {year} {2007})},\ \Eprint
  {http://arxiv.org/abs/astro-ph/0607207} {arXiv:astro-ph/0607207} \BibitemShut
  {NoStop}%
\bibitem [{\citenamefont {Calchi~Novati}\ \emph {et~al.}(2013)\citenamefont
  {Calchi~Novati}, \citenamefont {Mirzoyan}, \citenamefont {Jetzer},\ and\
  \citenamefont {Scarpetta}}]{calchinovati:microlensingconstraint}%
  \BibitemOpen
  \bibfield  {author} {\bibinfo {author} {\bibfnamefont {S.}~\bibnamefont
  {Calchi~Novati}}, \bibinfo {author} {\bibfnamefont {S.}~\bibnamefont
  {Mirzoyan}}, \bibinfo {author} {\bibfnamefont {P.}~\bibnamefont {Jetzer}}, \
  and\ \bibinfo {author} {\bibfnamefont {G.}~\bibnamefont {Scarpetta}},\ }\href
  {\doibase 10.1093/mnras/stt1402} {\bibfield  {journal} {\bibinfo  {journal}
  {MNRAS}\ }\textbf {\bibinfo {volume} {435}},\ \bibinfo {pages} {1582}
  (\bibinfo {year} {2013})},\ \Eprint {http://arxiv.org/abs/1308.4281}
  {arXiv:1308.4281} \BibitemShut {NoStop}%
\bibitem [{\citenamefont {Alcock}\ \emph {et~al.}(2001)\citenamefont {Alcock}
  \emph {et~al.}}]{alcock:microlensingconstraint}%
  \BibitemOpen
  \bibfield  {author} {\bibinfo {author} {\bibfnamefont {C.}~\bibnamefont
  {Alcock}} \emph {et~al.} (\bibinfo {collaboration} {MACHO Collaboration}),\
  }\href {\doibase 10.1086/319636} {\bibfield  {journal} {\bibinfo  {journal}
  {ApJ Letters}\ }\textbf {\bibinfo {volume} {550}},\ \bibinfo {pages} {L169}
  (\bibinfo {year} {2001})},\ \Eprint {http://arxiv.org/abs/astro-ph/0011506}
  {arXiv:astro-ph/0011506} \BibitemShut {NoStop}%
\bibitem [{\citenamefont {Mediavilla}\ \emph {et~al.}(2009)\citenamefont
  {Mediavilla}, \citenamefont {Munoz}, \citenamefont {Falco}, \citenamefont
  {Motta}, \citenamefont {Guerras}, \citenamefont {Canovas}, \citenamefont
  {Jean}, \citenamefont {Oscoz},\ and\ \citenamefont
  {Mosquera}}]{mediavilla:microlensingconstraint}%
  \BibitemOpen
  \bibfield  {author} {\bibinfo {author} {\bibfnamefont {E.}~\bibnamefont
  {Mediavilla}}, \bibinfo {author} {\bibfnamefont {J.~A.}\ \bibnamefont
  {Munoz}}, \bibinfo {author} {\bibfnamefont {E.}~\bibnamefont {Falco}},
  \bibinfo {author} {\bibfnamefont {V.}~\bibnamefont {Motta}}, \bibinfo
  {author} {\bibfnamefont {E.}~\bibnamefont {Guerras}}, \bibinfo {author}
  {\bibfnamefont {H.}~\bibnamefont {Canovas}}, \bibinfo {author} {\bibfnamefont
  {C.}~\bibnamefont {Jean}}, \bibinfo {author} {\bibfnamefont {A.}~\bibnamefont
  {Oscoz}}, \ and\ \bibinfo {author} {\bibfnamefont {A.~M.}\ \bibnamefont
  {Mosquera}},\ }\href {\doibase 10.1088/0004-637X/706/2/1451} {\bibfield
  {journal} {\bibinfo  {journal} {\apj}\ }\textbf {\bibinfo {volume} {706}},\
  \bibinfo {pages} {1451} (\bibinfo {year} {2009})},\ \Eprint
  {http://arxiv.org/abs/0910.3645} {arXiv:0910.3645} \BibitemShut {NoStop}%
\bibitem [{\citenamefont {Wilkinson}\ \emph {et~al.}(2001)\citenamefont
  {Wilkinson}, \citenamefont {Henstock}, \citenamefont {Browne}, \citenamefont
  {Polatidis}, \citenamefont {Augusto}, \citenamefont {Readhead}, \citenamefont
  {Pearson}, \citenamefont {Xu}, \citenamefont {Taylor},\ and\ \citenamefont
  {Vermeulen}}]{wilkinson:millilensingconstraint}%
  \BibitemOpen
  \bibfield  {author} {\bibinfo {author} {\bibfnamefont {P.~N.}\ \bibnamefont
  {Wilkinson}}, \bibinfo {author} {\bibfnamefont {D.~R.}\ \bibnamefont
  {Henstock}}, \bibinfo {author} {\bibfnamefont {I.~W.~A.}\ \bibnamefont
  {Browne}}, \bibinfo {author} {\bibfnamefont {A.~G.}\ \bibnamefont
  {Polatidis}}, \bibinfo {author} {\bibfnamefont {P.}~\bibnamefont {Augusto}},
  \bibinfo {author} {\bibfnamefont {A.~C.~S.}\ \bibnamefont {Readhead}},
  \bibinfo {author} {\bibfnamefont {T.~J.}\ \bibnamefont {Pearson}}, \bibinfo
  {author} {\bibfnamefont {W.}~\bibnamefont {Xu}}, \bibinfo {author}
  {\bibfnamefont {G.~B.}\ \bibnamefont {Taylor}}, \ and\ \bibinfo {author}
  {\bibfnamefont {R.~C.}\ \bibnamefont {Vermeulen}},\ }\href {\doibase
  10.1103/PhysRevLett.86.584} {\bibfield  {journal} {\bibinfo  {journal} {Phys.
  Rev. Lett.}\ }\textbf {\bibinfo {volume} {86}},\ \bibinfo {pages} {584}
  (\bibinfo {year} {2001})},\ \Eprint {http://arxiv.org/abs/astro-ph/0101328}
  {arXiv:astro-ph/0101328} \BibitemShut {NoStop}%
\bibitem [{\citenamefont {Zumalacarregui}\ and\ \citenamefont
  {Seljak}(2018)}]{zumalacarregui:supernovaconstraint}%
  \BibitemOpen
  \bibfield  {author} {\bibinfo {author} {\bibfnamefont {M.}~\bibnamefont
  {Zumalacarregui}}\ and\ \bibinfo {author} {\bibfnamefont {U.}~\bibnamefont
  {Seljak}},\ }\href {\doibase 10.1103/PhysRevLett.121.141101} {\bibfield
  {journal} {\bibinfo  {journal} {Phys. Rev. Lett.}\ }\textbf {\bibinfo
  {volume} {121}},\ \bibinfo {pages} {141101} (\bibinfo {year} {2018})},\
  \Eprint {http://arxiv.org/abs/1712.02240} {arXiv:1712.02240 [astro-ph.CO]}
  \BibitemShut {NoStop}%
%%CITATION = ARXIV:1712.02240;%%
\bibitem [{\citenamefont {Graham}\ \emph {et~al.}(2015)\citenamefont {Graham},
  \citenamefont {Rajendran},\ and\ \citenamefont
  {Varela}}]{graham:whitedwarfconstraint}%
  \BibitemOpen
  \bibfield  {author} {\bibinfo {author} {\bibfnamefont {P.~W.}\ \bibnamefont
  {Graham}}, \bibinfo {author} {\bibfnamefont {S.}~\bibnamefont {Rajendran}}, \
  and\ \bibinfo {author} {\bibfnamefont {J.}~\bibnamefont {Varela}},\ }\href
  {\doibase 10.1103/PhysRevD.92.063007} {\bibfield  {journal} {\bibinfo
  {journal} {Phys. Rev. D}\ }\textbf {\bibinfo {volume} {92}},\ \bibinfo
  {pages} {063007} (\bibinfo {year} {2015})},\ \Eprint
  {http://arxiv.org/abs/1505.04444} {arXiv:1505.04444} \BibitemShut {NoStop}%
\bibitem [{\citenamefont {Capela}\ \emph {et~al.}(2013)\citenamefont {Capela},
  \citenamefont {Pshirkov},\ and\ \citenamefont
  {Tinyakov}}]{capela:neutronstarconstaint}%
  \BibitemOpen
  \bibfield  {author} {\bibinfo {author} {\bibfnamefont {F.}~\bibnamefont
  {Capela}}, \bibinfo {author} {\bibfnamefont {M.}~\bibnamefont {Pshirkov}}, \
  and\ \bibinfo {author} {\bibfnamefont {P.}~\bibnamefont {Tinyakov}},\ }\href
  {\doibase 10.1103/PhysRevD.87.123524} {\bibfield  {journal} {\bibinfo
  {journal} {Phys. Rev. D}\ }\textbf {\bibinfo {volume} {87}},\ \bibinfo
  {pages} {123524} (\bibinfo {year} {2013})},\ \Eprint
  {http://arxiv.org/abs/1301.4984} {arXiv:1301.4984} \BibitemShut {NoStop}%
\bibitem [{\citenamefont {Quinn}\ \emph {et~al.}(2009)\citenamefont {Quinn},
  \citenamefont {Wilkinson}, \citenamefont {Irwin}, \citenamefont {Marshall},
  \citenamefont {Koch},\ and\ \citenamefont
  {Belokurov}}]{quinn:widebinaryconstraint}%
  \BibitemOpen
  \bibfield  {author} {\bibinfo {author} {\bibfnamefont {D.~P.}\ \bibnamefont
  {Quinn}}, \bibinfo {author} {\bibfnamefont {M.~I.}\ \bibnamefont
  {Wilkinson}}, \bibinfo {author} {\bibfnamefont {M.~J.}\ \bibnamefont
  {Irwin}}, \bibinfo {author} {\bibfnamefont {J.}~\bibnamefont {Marshall}},
  \bibinfo {author} {\bibfnamefont {A.}~\bibnamefont {Koch}}, \ and\ \bibinfo
  {author} {\bibfnamefont {V.}~\bibnamefont {Belokurov}},\ }\href {\doibase
  10.1111/j.1745-3933.2009.00652.x} {\bibfield  {journal} {\bibinfo  {journal}
  {MNRAS Letters}\ }\textbf {\bibinfo {volume} {396}},\ \bibinfo {pages} {L11}
  (\bibinfo {year} {2009})},\ \Eprint {http://arxiv.org/abs/0903.1644}
  {arXiv:0903.1644} \BibitemShut {NoStop}%
\bibitem [{\citenamefont {Brandt}(2016)}]{brandt:ufdgconstraint}%
  \BibitemOpen
  \bibfield  {author} {\bibinfo {author} {\bibfnamefont {T.~D.}\ \bibnamefont
  {Brandt}},\ }\href {\doibase 10.3847/2041-8205/824/2/L31} {\bibfield
  {journal} {\bibinfo  {journal} {ApJ Letters}\ }\textbf {\bibinfo {volume}
  {824}},\ \bibinfo {pages} {L31} (\bibinfo {year} {2016})},\ \Eprint
  {http://arxiv.org/abs/1605.03665} {arXiv:1605.03665} \BibitemShut {NoStop}%
\bibitem [{\citenamefont {Raidal}\ \emph {et~al.}(2017)\citenamefont {Raidal},
  \citenamefont {Vaskonen},\ and\ \citenamefont {Veermäe}}]{Raidal:2017mfl}%
  \BibitemOpen
  \bibfield  {author} {\bibinfo {author} {\bibfnamefont {M.}~\bibnamefont
  {Raidal}}, \bibinfo {author} {\bibfnamefont {V.}~\bibnamefont {Vaskonen}}, \
  and\ \bibinfo {author} {\bibfnamefont {H.}~\bibnamefont {Veermäe}},\ }\href
  {\doibase 10.1088/1475-7516/2017/09/037} {\bibfield  {journal} {\bibinfo
  {journal} {JCAP}\ }\textbf {\bibinfo {volume} {1709}},\ \bibinfo {pages}
  {037} (\bibinfo {year} {2017})},\ \Eprint {http://arxiv.org/abs/1707.01480}
  {arXiv:1707.01480 [astro-ph.CO]} \BibitemShut {NoStop}%
%%CITATION = ARXIV:1707.01480;%%
\bibitem [{\citenamefont {Ali-Ha\"{\i}moud}\ \emph {et~al.}(2017)\citenamefont
  {Ali-Ha\"{\i}moud}, \citenamefont {Kovetz},\ and\ \citenamefont
  {Kamionkowski}}]{alihaimoud:pbhmergerrate}%
  \BibitemOpen
  \bibfield  {author} {\bibinfo {author} {\bibfnamefont {Y.}~\bibnamefont
  {Ali-Ha\"{\i}moud}}, \bibinfo {author} {\bibfnamefont {E.~D.}\ \bibnamefont
  {Kovetz}}, \ and\ \bibinfo {author} {\bibfnamefont {M.}~\bibnamefont
  {Kamionkowski}},\ }\href {\doibase 10.1103/PhysRevD.96.123523} {\bibfield
  {journal} {\bibinfo  {journal} {Phys. Rev. D}\ }\textbf {\bibinfo {volume}
  {96}},\ \bibinfo {pages} {123523} (\bibinfo {year} {2017})},\ \Eprint
  {http://arxiv.org/abs/1709.06576} {arXiv:1709.06576} \BibitemShut {NoStop}%
\bibitem [{\citenamefont {Raidal}\ \emph {et~al.}(2019)\citenamefont {Raidal},
  \citenamefont {Spethmann}, \citenamefont {Vaskonen},\ and\ \citenamefont
  {Veermäe}}]{Raidal:2018bbj}%
  \BibitemOpen
  \bibfield  {author} {\bibinfo {author} {\bibfnamefont {M.}~\bibnamefont
  {Raidal}}, \bibinfo {author} {\bibfnamefont {C.}~\bibnamefont {Spethmann}},
  \bibinfo {author} {\bibfnamefont {V.}~\bibnamefont {Vaskonen}}, \ and\
  \bibinfo {author} {\bibfnamefont {H.}~\bibnamefont {Veermäe}},\ }\href
  {\doibase 10.1088/1475-7516/2019/02/018} {\bibfield  {journal} {\bibinfo
  {journal} {JCAP}\ }\textbf {\bibinfo {volume} {1902}},\ \bibinfo {pages}
  {018} (\bibinfo {year} {2019})},\ \Eprint {http://arxiv.org/abs/1812.01930}
  {arXiv:1812.01930 [astro-ph.CO]} \BibitemShut {NoStop}%
%%CITATION = ARXIV:1812.01930;%%
\bibitem [{\citenamefont {Magee}\ \emph {et~al.}(2018)\citenamefont {Magee},
  \citenamefont {Deutsch}, \citenamefont {McClincy}, \citenamefont {Hanna},
  \citenamefont {Horst}, \citenamefont {Meacher}, \citenamefont {Messick},
  \citenamefont {Shandera},\ and\ \citenamefont {Wade}}]{magee:mergerrate}%
  \BibitemOpen
  \bibfield  {author} {\bibinfo {author} {\bibfnamefont {R.}~\bibnamefont
  {Magee}}, \bibinfo {author} {\bibfnamefont {A.-S.}\ \bibnamefont {Deutsch}},
  \bibinfo {author} {\bibfnamefont {P.}~\bibnamefont {McClincy}}, \bibinfo
  {author} {\bibfnamefont {C.}~\bibnamefont {Hanna}}, \bibinfo {author}
  {\bibfnamefont {C.}~\bibnamefont {Horst}}, \bibinfo {author} {\bibfnamefont
  {D.}~\bibnamefont {Meacher}}, \bibinfo {author} {\bibfnamefont
  {C.}~\bibnamefont {Messick}}, \bibinfo {author} {\bibfnamefont
  {S.}~\bibnamefont {Shandera}}, \ and\ \bibinfo {author} {\bibfnamefont
  {M.}~\bibnamefont {Wade}},\ }\href {\doibase 10.1103/PhysRevD.98.103024}
  {\bibfield  {journal} {\bibinfo  {journal} {Phys. Rev. D}\ }\textbf {\bibinfo
  {volume} {98}},\ \bibinfo {pages} {103024} (\bibinfo {year} {2018})},\
  \Eprint {http://arxiv.org/abs/1808.04772} {arXiv:1808.04772 [astro-ph.IM]}
  \BibitemShut {NoStop}%
%%CITATION = ARXIV:1808.04772;%%
\bibitem [{\citenamefont {Gaggero}\ \emph {et~al.}(2017)\citenamefont
  {Gaggero}, \citenamefont {Bertone}, \citenamefont {Calore}, \citenamefont
  {Connors}, \citenamefont {Lovell}, \citenamefont {Markoff},\ and\
  \citenamefont {Storm}}]{gaggero:accretionconstraints}%
  \BibitemOpen
  \bibfield  {author} {\bibinfo {author} {\bibfnamefont {D.}~\bibnamefont
  {Gaggero}}, \bibinfo {author} {\bibfnamefont {G.}~\bibnamefont {Bertone}},
  \bibinfo {author} {\bibfnamefont {F.}~\bibnamefont {Calore}}, \bibinfo
  {author} {\bibfnamefont {R.~M.~T.}\ \bibnamefont {Connors}}, \bibinfo
  {author} {\bibfnamefont {M.}~\bibnamefont {Lovell}}, \bibinfo {author}
  {\bibfnamefont {S.}~\bibnamefont {Markoff}}, \ and\ \bibinfo {author}
  {\bibfnamefont {E.}~\bibnamefont {Storm}},\ }\href {\doibase
  10.1103/PhysRevLett.118.241101} {\bibfield  {journal} {\bibinfo  {journal}
  {Phys. Rev. Lett.}\ }\textbf {\bibinfo {volume} {118}},\ \bibinfo {pages}
  {241101} (\bibinfo {year} {2017})},\ \Eprint
  {http://arxiv.org/abs/1612.00457} {arXiv:1612.00457} \BibitemShut {NoStop}%
\bibitem [{\citenamefont {Ricotti}\ \emph {et~al.}(2008)\citenamefont
  {Ricotti}, \citenamefont {Ostriker},\ and\ \citenamefont
  {Mack}}]{ricotti:cmbconstraint}%
  \BibitemOpen
  \bibfield  {author} {\bibinfo {author} {\bibfnamefont {M.}~\bibnamefont
  {Ricotti}}, \bibinfo {author} {\bibfnamefont {J.~P.}\ \bibnamefont
  {Ostriker}}, \ and\ \bibinfo {author} {\bibfnamefont {K.~J.}\ \bibnamefont
  {Mack}},\ }\href {\doibase 10.1086/587831} {\bibfield  {journal} {\bibinfo
  {journal} {\apj}\ }\textbf {\bibinfo {volume} {680}},\ \bibinfo {pages} {829}
  (\bibinfo {year} {2008})},\ \Eprint {http://arxiv.org/abs/0709.0524}
  {arXiv:0709.0524} \BibitemShut {NoStop}%
\bibitem [{\citenamefont {Ali-Ha\"{\i}moud}\ and\ \citenamefont
  {Kamionkowski}(2017)}]{alihamoud:pbhaccretion}%
  \BibitemOpen
  \bibfield  {author} {\bibinfo {author} {\bibfnamefont {Y.}~\bibnamefont
  {Ali-Ha\"{\i}moud}}\ and\ \bibinfo {author} {\bibfnamefont {M.}~\bibnamefont
  {Kamionkowski}},\ }\href {\doibase 10.1103/PhysRevD.95.043534} {\bibfield
  {journal} {\bibinfo  {journal} {Phys. Rev. D}\ }\textbf {\bibinfo {volume}
  {95}},\ \bibinfo {pages} {043534} (\bibinfo {year} {2017})},\ \Eprint
  {http://arxiv.org/abs/1612.05644} {arXiv:1612.05644} \BibitemShut {NoStop}%
\bibitem [{\citenamefont {Poulin}\ \emph {et~al.}(2017)\citenamefont {Poulin},
  \citenamefont {Serpico}, \citenamefont {Calore}, \citenamefont {Clesse},\
  and\ \citenamefont {Kohri}}]{poulin:cmbconstraint}%
  \BibitemOpen
  \bibfield  {author} {\bibinfo {author} {\bibfnamefont {V.}~\bibnamefont
  {Poulin}}, \bibinfo {author} {\bibfnamefont {P.~D.}\ \bibnamefont {Serpico}},
  \bibinfo {author} {\bibfnamefont {F.}~\bibnamefont {Calore}}, \bibinfo
  {author} {\bibfnamefont {S.}~\bibnamefont {Clesse}}, \ and\ \bibinfo {author}
  {\bibfnamefont {K.}~\bibnamefont {Kohri}},\ }\href {\doibase
  10.1103/PhysRevD.96.083524} {\bibfield  {journal} {\bibinfo  {journal} {Phys.
  Rev. D}\ }\textbf {\bibinfo {volume} {96}},\ \bibinfo {pages} {083524}
  (\bibinfo {year} {2017})},\ \Eprint {http://arxiv.org/abs/1707.04206}
  {arXiv:1707.04206} \BibitemShut {NoStop}%
\bibitem [{\citenamefont {Bernal}\ \emph {et~al.}(2017)\citenamefont {Bernal},
  \citenamefont {Bellomo}, \citenamefont {Raccanelli},\ and\ \citenamefont
  {Verde}}]{bernal:cmbconstraint}%
  \BibitemOpen
  \bibfield  {author} {\bibinfo {author} {\bibfnamefont {J.~L.}\ \bibnamefont
  {Bernal}}, \bibinfo {author} {\bibfnamefont {N.}~\bibnamefont {Bellomo}},
  \bibinfo {author} {\bibfnamefont {A.}~\bibnamefont {Raccanelli}}, \ and\
  \bibinfo {author} {\bibfnamefont {L.}~\bibnamefont {Verde}},\ }\href
  {\doibase 10.1088/1475-7516/2017/10/052} {\bibfield  {journal} {\bibinfo
  {journal} {JCAP}\ }\textbf {\bibinfo {volume} {2017}},\ \bibinfo {pages}
  {052} (\bibinfo {year} {2017})},\ \Eprint {http://arxiv.org/abs/1709.07465}
  {arXiv:1709.07465} \BibitemShut {NoStop}%
\bibitem [{\citenamefont {Aloni}\ \emph {et~al.}(2017)\citenamefont {Aloni},
  \citenamefont {Blum},\ and\ \citenamefont {Flauger}}]{Aloni2017}%
  \BibitemOpen
  \bibfield  {author} {\bibinfo {author} {\bibfnamefont {D.}~\bibnamefont
  {Aloni}}, \bibinfo {author} {\bibfnamefont {K.}~\bibnamefont {Blum}}, \ and\
  \bibinfo {author} {\bibfnamefont {R.}~\bibnamefont {Flauger}},\ }\href
  {\doibase 10.1088/1475-7516/2017/05/017} {\bibfield  {journal} {\bibinfo
  {journal} {JCAP}\ }\textbf {\bibinfo {volume} {2017}},\ \bibinfo {pages}
  {017} (\bibinfo {year} {2017})},\ \Eprint {http://arxiv.org/abs/1612.06811}
  {arXiv:1612.06811} \BibitemShut {NoStop}%
\bibitem [{\citenamefont {Bellomo}\ \emph {et~al.}(2018)\citenamefont
  {Bellomo}, \citenamefont {Bernal}, \citenamefont {Raccanelli},\ and\
  \citenamefont {Verde}}]{bellomo:emdconstraints}%
  \BibitemOpen
  \bibfield  {author} {\bibinfo {author} {\bibfnamefont {N.}~\bibnamefont
  {Bellomo}}, \bibinfo {author} {\bibfnamefont {J.~L.}\ \bibnamefont {Bernal}},
  \bibinfo {author} {\bibfnamefont {A.}~\bibnamefont {Raccanelli}}, \ and\
  \bibinfo {author} {\bibfnamefont {L.}~\bibnamefont {Verde}},\ }\href
  {\doibase 10.1088/1475-7516/2018/01/004} {\bibfield  {journal} {\bibinfo
  {journal} {JCAP}\ }\textbf {\bibinfo {volume} {2018}},\ \bibinfo {pages}
  {004} (\bibinfo {year} {2018})},\ \Eprint {http://arxiv.org/abs/1709.07467}
  {arXiv:1709.07467} \BibitemShut {NoStop}%
\bibitem [{\citenamefont {Nakama}\ \emph {et~al.}(2017)\citenamefont {Nakama},
  \citenamefont {Silk},\ and\ \citenamefont {Kamionkowski}}]{Nakama:NG}%
  \BibitemOpen
  \bibfield  {author} {\bibinfo {author} {\bibfnamefont {T.}~\bibnamefont
  {Nakama}}, \bibinfo {author} {\bibfnamefont {J.}~\bibnamefont {Silk}}, \ and\
  \bibinfo {author} {\bibfnamefont {M.}~\bibnamefont {Kamionkowski}},\ }\href
  {\doibase 10.1103/PhysRevD.95.043511} {\bibfield  {journal} {\bibinfo
  {journal} {Phys. Rev. D}\ }\textbf {\bibinfo {volume} {95}},\ \bibinfo
  {pages} {043511} (\bibinfo {year} {2017})}\BibitemShut {NoStop}%
\bibitem [{\citenamefont {Sasaki}\ \emph {et~al.}(2018)\citenamefont {Sasaki},
  \citenamefont {Suyama}, \citenamefont {Tanaka},\ and\ \citenamefont
  {Yokoyama}}]{Sasaki_2018}%
  \BibitemOpen
  \bibfield  {author} {\bibinfo {author} {\bibfnamefont {M.}~\bibnamefont
  {Sasaki}}, \bibinfo {author} {\bibfnamefont {T.}~\bibnamefont {Suyama}},
  \bibinfo {author} {\bibfnamefont {T.}~\bibnamefont {Tanaka}}, \ and\ \bibinfo
  {author} {\bibfnamefont {S.}~\bibnamefont {Yokoyama}},\ }\href {\doibase
  10.1088/1361-6382/aaa7b4} {\bibfield  {journal} {\bibinfo  {journal}
  {Classical and Quantum Gravity}\ }\textbf {\bibinfo {volume} {35}},\ \bibinfo
  {pages} {063001} (\bibinfo {year} {2018})},\ \Eprint
  {http://arxiv.org/abs/1801.05235} {arXiv:1801.05235} \BibitemShut {NoStop}%
\bibitem [{\citenamefont {Carr}\ and\ \citenamefont
  {Silk}(2018)}]{Carr_Silk_2018}%
  \BibitemOpen
  \bibfield  {author} {\bibinfo {author} {\bibfnamefont {B.}~\bibnamefont
  {Carr}}\ and\ \bibinfo {author} {\bibfnamefont {J.}~\bibnamefont {Silk}},\
  }\href {\doibase 10.1093/mnras/sty1204} {\bibfield  {journal} {\bibinfo
  {journal} {MNRAS}\ }\textbf {\bibinfo {volume} {478}},\ \bibinfo {pages}
  {3756} (\bibinfo {year} {2018})},\ \Eprint {http://arxiv.org/abs/1801.00672}
  {arXiv:1801.00672} \BibitemShut {NoStop}%
\bibitem [{\citenamefont {Ali-Haimoud}\ \emph {et~al.}(2019)\citenamefont
  {Ali-Haimoud} \emph {et~al.}}]{Ali-Haimoud:2019khd}%
  \BibitemOpen
  \bibfield  {author} {\bibinfo {author} {\bibfnamefont {Y.}~\bibnamefont
  {Ali-Haimoud}} \emph {et~al.},\ }\href@noop {} {\  (\bibinfo {year}
  {2019})},\ \Eprint {http://arxiv.org/abs/1903.04424} {arXiv:1903.04424
  [astro-ph.CO]} \BibitemShut {NoStop}%
%%CITATION = ARXIV:1903.04424;%%
\bibitem [{\citenamefont {Viel}\ \emph {et~al.}(2002)\citenamefont {Viel},
  \citenamefont {Matarrese}, \citenamefont {Mo}, \citenamefont {Haehnelt},\
  and\ \citenamefont {Theuns}}]{Viel:2001hd}%
  \BibitemOpen
  \bibfield  {author} {\bibinfo {author} {\bibfnamefont {M.}~\bibnamefont
  {Viel}}, \bibinfo {author} {\bibfnamefont {S.}~\bibnamefont {Matarrese}},
  \bibinfo {author} {\bibfnamefont {H.~J.}\ \bibnamefont {Mo}}, \bibinfo
  {author} {\bibfnamefont {M.~G.}\ \bibnamefont {Haehnelt}}, \ and\ \bibinfo
  {author} {\bibfnamefont {T.}~\bibnamefont {Theuns}},\ }\href {\doibase
  10.1046/j.1365-8711.2002.05060.x} {\bibfield  {journal} {\bibinfo  {journal}
  {MNRAS}\ }\textbf {\bibinfo {volume} {329}},\ \bibinfo {pages} {848}
  (\bibinfo {year} {2002})},\ \Eprint {http://arxiv.org/abs/astro-ph/0105233}
  {arXiv:astro-ph/0105233 [astro-ph]} \BibitemShut {NoStop}%
%%CITATION = ASTRO-PH/0105233;%%
\bibitem [{\citenamefont {Viel}\ \emph {et~al.}(2005)\citenamefont {Viel},
  \citenamefont {Lesgourgues}, \citenamefont {Haehnelt}, \citenamefont
  {Matarrese},\ and\ \citenamefont {Riotto}}]{Viel2005}%
  \BibitemOpen
  \bibfield  {author} {\bibinfo {author} {\bibfnamefont {M.}~\bibnamefont
  {Viel}}, \bibinfo {author} {\bibfnamefont {J.}~\bibnamefont {Lesgourgues}},
  \bibinfo {author} {\bibfnamefont {M.~G.}\ \bibnamefont {Haehnelt}}, \bibinfo
  {author} {\bibfnamefont {S.}~\bibnamefont {Matarrese}}, \ and\ \bibinfo
  {author} {\bibfnamefont {A.}~\bibnamefont {Riotto}},\ }\href {\doibase
  10.1103/PhysRevD.71.063534} {\bibfield  {journal} {\bibinfo  {journal} {Phys.
  Rev.}\ }\textbf {\bibinfo {volume} {D71}},\ \bibinfo {pages} {063534}
  (\bibinfo {year} {2005})},\ \Eprint {http://arxiv.org/abs/astro-ph/0501562}
  {arXiv:astro-ph/0501562 [astro-ph]} \BibitemShut {NoStop}%
%%CITATION = ASTRO-PH/0501562;%%
\bibitem [{\citenamefont {Viel}\ \emph {et~al.}(2013)\citenamefont {Viel},
  \citenamefont {Becker}, \citenamefont {Bolton},\ and\ \citenamefont
  {Haehnelt}}]{Viel:2013apy}%
  \BibitemOpen
  \bibfield  {author} {\bibinfo {author} {\bibfnamefont {M.}~\bibnamefont
  {Viel}}, \bibinfo {author} {\bibfnamefont {G.~D.}\ \bibnamefont {Becker}},
  \bibinfo {author} {\bibfnamefont {J.~S.}\ \bibnamefont {Bolton}}, \ and\
  \bibinfo {author} {\bibfnamefont {M.~G.}\ \bibnamefont {Haehnelt}},\ }\href
  {\doibase 10.1103/PhysRevD.88.043502} {\bibfield  {journal} {\bibinfo
  {journal} {Phys. Rev. D}\ }\textbf {\bibinfo {volume} {88}},\ \bibinfo
  {pages} {043502} (\bibinfo {year} {2013})},\ \Eprint
  {http://arxiv.org/abs/1306.2314} {arXiv:1306.2314 [astro-ph.CO]} \BibitemShut
  {NoStop}%
%%CITATION = ARXIV:1306.2314;%%
\bibitem [{\citenamefont {Afshordi}\ \emph {et~al.}(2003)\citenamefont
  {Afshordi}, \citenamefont {McDonald},\ and\ \citenamefont
  {Spergel}}]{Afshordi:2003zb}%
  \BibitemOpen
  \bibfield  {author} {\bibinfo {author} {\bibfnamefont {N.}~\bibnamefont
  {Afshordi}}, \bibinfo {author} {\bibfnamefont {P.}~\bibnamefont {McDonald}},
  \ and\ \bibinfo {author} {\bibfnamefont {D.~N.}\ \bibnamefont {Spergel}},\
  }\href {\doibase 10.1086/378763} {\bibfield  {journal} {\bibinfo  {journal}
  {\apj}\ }\textbf {\bibinfo {volume} {594}},\ \bibinfo {pages} {L71} (\bibinfo
  {year} {2003})},\ \Eprint {http://arxiv.org/abs/astro-ph/0302035}
  {arXiv:astro-ph/0302035 [astro-ph]} \BibitemShut {NoStop}%
%%CITATION = ASTRO-PH/0302035;%%
\bibitem [{\citenamefont {Meszaros}(1975)}]{meszaros:primeval}%
  \BibitemOpen
  \bibfield  {author} {\bibinfo {author} {\bibfnamefont {P.}~\bibnamefont
  {Meszaros}},\ }\href@noop {} {\bibfield  {journal} {\bibinfo  {journal}
  {Astronomy and Astrophysics}\ }\textbf {\bibinfo {volume} {38}},\ \bibinfo
  {pages} {5} (\bibinfo {year} {1975})}\BibitemShut {NoStop}%
\bibitem [{\citenamefont {Gong}\ and\ \citenamefont
  {Kitajima}(2017)}]{Gong:2017sie}%
  \BibitemOpen
  \bibfield  {author} {\bibinfo {author} {\bibfnamefont {J.-O.}\ \bibnamefont
  {Gong}}\ and\ \bibinfo {author} {\bibfnamefont {N.}~\bibnamefont
  {Kitajima}},\ }\href {\doibase 10.1088/1475-7516/2017/08/017} {\bibfield
  {journal} {\bibinfo  {journal} {JCAP}\ }\textbf {\bibinfo {volume} {1708}},\
  \bibinfo {pages} {017} (\bibinfo {year} {2017})},\ \Eprint
  {http://arxiv.org/abs/1704.04132} {arXiv:1704.04132 [astro-ph.CO]}
  \BibitemShut {NoStop}%
%%CITATION = ARXIV:1704.04132;%%
\bibitem [{\citenamefont {Kalaja}\ \emph {et~al.}(tion)\citenamefont {Kalaja}
  \emph {et~al.}}]{Kalaja}%
  \BibitemOpen
  \bibfield  {author} {\bibinfo {author} {\bibfnamefont {A.}~\bibnamefont
  {Kalaja}} \emph {et~al.},\ }\href@noop {} {\  (\bibinfo {year} {in
  preparation})}\BibitemShut {NoStop}%
\bibitem [{\citenamefont {Carr}\ \emph {et~al.}(2017)\citenamefont {Carr},
  \citenamefont {Raidal}, \citenamefont {Tenkanen}, \citenamefont {Vaskonen},\
  and\ \citenamefont {Veermae}}]{Carr:2017jsz}%
  \BibitemOpen
  \bibfield  {author} {\bibinfo {author} {\bibfnamefont {B.}~\bibnamefont
  {Carr}}, \bibinfo {author} {\bibfnamefont {M.}~\bibnamefont {Raidal}},
  \bibinfo {author} {\bibfnamefont {T.}~\bibnamefont {Tenkanen}}, \bibinfo
  {author} {\bibfnamefont {V.}~\bibnamefont {Vaskonen}}, \ and\ \bibinfo
  {author} {\bibfnamefont {H.}~\bibnamefont {Veermae}},\ }\href {\doibase
  10.1103/PhysRevD.96.023514} {\bibfield  {journal} {\bibinfo  {journal} {Phys.
  Rev.}\ }\textbf {\bibinfo {volume} {D96}},\ \bibinfo {pages} {023514}
  (\bibinfo {year} {2017})},\ \Eprint {http://arxiv.org/abs/1705.05567}
  {arXiv:1705.05567 [astro-ph.CO]} \BibitemShut {NoStop}%
%%CITATION = ARXIV:1705.05567;%%
\bibitem [{\citenamefont {Dolgov}\ and\ \citenamefont
  {Silk}(1993)}]{dolgov_silk}%
  \BibitemOpen
  \bibfield  {author} {\bibinfo {author} {\bibfnamefont {A.}~\bibnamefont
  {Dolgov}}\ and\ \bibinfo {author} {\bibfnamefont {J.}~\bibnamefont {Silk}},\
  }\href {\doibase 10.1103/PhysRevD.47.4244} {\bibfield  {journal} {\bibinfo
  {journal} {Phys. Rev. D}\ }\textbf {\bibinfo {volume} {47}},\ \bibinfo
  {pages} {4244} (\bibinfo {year} {1993})}\BibitemShut {NoStop}%
\bibitem [{\citenamefont {Green}(2016)}]{green:2016}%
  \BibitemOpen
  \bibfield  {author} {\bibinfo {author} {\bibfnamefont {A.~M.}\ \bibnamefont
  {Green}},\ }\href {\doibase 10.1103/PhysRevD.94.063530} {\bibfield  {journal}
  {\bibinfo  {journal} {Phys. Rev. D}\ }\textbf {\bibinfo {volume} {94}},\
  \bibinfo {pages} {063530} (\bibinfo {year} {2016})}\BibitemShut {NoStop}%
\bibitem [{\citenamefont {Kannike}\ \emph {et~al.}(2017)\citenamefont
  {Kannike}, \citenamefont {Marzola}, \citenamefont {Raidal},\ and\
  \citenamefont {Veerm{\"a}e}}]{kannike:2017}%
  \BibitemOpen
  \bibfield  {author} {\bibinfo {author} {\bibfnamefont {K.}~\bibnamefont
  {Kannike}}, \bibinfo {author} {\bibfnamefont {L.}~\bibnamefont {Marzola}},
  \bibinfo {author} {\bibfnamefont {M.}~\bibnamefont {Raidal}}, \ and\ \bibinfo
  {author} {\bibfnamefont {H.}~\bibnamefont {Veerm{\"a}e}},\ }\href {\doibase
  10.1088/1475-7516/2017/09/020} {\bibfield  {journal} {\bibinfo  {journal}
  {JCAP}\ }\textbf {\bibinfo {volume} {2017}},\ \bibinfo {pages} {020}
  (\bibinfo {year} {2017})}\BibitemShut {NoStop}%
\bibitem [{\citenamefont {{Carr}}(1975)}]{carr:1975primordialmassspectrum}%
  \BibitemOpen
  \bibfield  {author} {\bibinfo {author} {\bibfnamefont {B.~J.}\ \bibnamefont
  {{Carr}}},\ }\href {\doibase 10.1086/153853} {\bibfield  {journal} {\bibinfo
  {journal} {\apj}\ }\textbf {\bibinfo {volume} {201}},\ \bibinfo {pages} {1}
  (\bibinfo {year} {1975})}\BibitemShut {NoStop}%
\bibitem [{\citenamefont {Murgia}\ \emph {et~al.}(2018)\citenamefont {Murgia},
  \citenamefont {Ir\v{s}i\v{c}},\ and\ \citenamefont {Viel}}]{Murgia:2018now}%
  \BibitemOpen
  \bibfield  {author} {\bibinfo {author} {\bibfnamefont {R.}~\bibnamefont
  {Murgia}}, \bibinfo {author} {\bibfnamefont {V.}~\bibnamefont
  {Ir\v{s}i\v{c}}}, \ and\ \bibinfo {author} {\bibfnamefont {M.}~\bibnamefont
  {Viel}},\ }\href {\doibase 10.1103/PhysRevD.98.083540} {\bibfield  {journal}
  {\bibinfo  {journal} {Phys. Rev. D}\ }\textbf {\bibinfo {volume} {98}},\
  \bibinfo {pages} {083540} (\bibinfo {year} {2018})},\ \Eprint
  {http://arxiv.org/abs/1806.08371} {arXiv:1806.08371 [astro-ph.CO]}
  \BibitemShut {NoStop}%
%%CITATION = ARXIV:1806.08371;%%
\bibitem [{\citenamefont {Springel}\ \emph {et~al.}(2001)\citenamefont
  {Springel}, \citenamefont {Yoshida},\ and\ \citenamefont
  {White}}]{Springel:2000yr}%
  \BibitemOpen
  \bibfield  {author} {\bibinfo {author} {\bibfnamefont {V.}~\bibnamefont
  {Springel}}, \bibinfo {author} {\bibfnamefont {N.}~\bibnamefont {Yoshida}}, \
  and\ \bibinfo {author} {\bibfnamefont {S.~D.~M.}\ \bibnamefont {White}},\
  }\href {\doibase 10.1016/S1384-1076(01)00042-2} {\bibfield  {journal}
  {\bibinfo  {journal} {New Astron.}\ }\textbf {\bibinfo {volume} {6}},\
  \bibinfo {pages} {79} (\bibinfo {year} {2001})},\ \Eprint
  {http://arxiv.org/abs/astro-ph/0003162} {arXiv:astro-ph/0003162 [astro-ph]}
  \BibitemShut {NoStop}%
%%CITATION = ASTRO-PH/0003162;%%
\bibitem [{\citenamefont {Springel}(2005)}]{Springel:2005mi}%
  \BibitemOpen
  \bibfield  {author} {\bibinfo {author} {\bibfnamefont {V.}~\bibnamefont
  {Springel}},\ }\href {\doibase 10.1111/j.1365-2966.2005.09655.x} {\bibfield
  {journal} {\bibinfo  {journal} {MNRAS}\ }\textbf {\bibinfo {volume} {364}},\
  \bibinfo {pages} {1105} (\bibinfo {year} {2005})},\ \Eprint
  {http://arxiv.org/abs/astro-ph/0505010} {arXiv:astro-ph/0505010 [astro-ph]}
  \BibitemShut {NoStop}%
%%CITATION = ASTRO-PH/0505010;%%
\bibitem [{\citenamefont {Crocce}\ \emph {et~al.}(2006)\citenamefont {Crocce},
  \citenamefont {Pueblas},\ and\ \citenamefont {Scoccimarro}}]{Crocce:2006ve}%
  \BibitemOpen
  \bibfield  {author} {\bibinfo {author} {\bibfnamefont {M.}~\bibnamefont
  {Crocce}}, \bibinfo {author} {\bibfnamefont {S.}~\bibnamefont {Pueblas}}, \
  and\ \bibinfo {author} {\bibfnamefont {R.}~\bibnamefont {Scoccimarro}},\
  }\href {\doibase 10.1111/j.1365-2966.2006.11040.x} {\bibfield  {journal}
  {\bibinfo  {journal} {MNRAS}\ }\textbf {\bibinfo {volume} {373}},\ \bibinfo
  {pages} {369} (\bibinfo {year} {2006})},\ \Eprint
  {http://arxiv.org/abs/astro-ph/0606505} {arXiv:astro-ph/0606505 [astro-ph]}
  \BibitemShut {NoStop}%
%%CITATION = ASTRO-PH/0606505;%%
\bibitem [{\citenamefont {Blas}\ \emph {et~al.}(2011)\citenamefont {Blas},
  \citenamefont {Lesgourgues},\ and\ \citenamefont {Tram}}]{Blas:2011rf}%
  \BibitemOpen
  \bibfield  {author} {\bibinfo {author} {\bibfnamefont {D.}~\bibnamefont
  {Blas}}, \bibinfo {author} {\bibfnamefont {J.}~\bibnamefont {Lesgourgues}}, \
  and\ \bibinfo {author} {\bibfnamefont {T.}~\bibnamefont {Tram}},\ }\href
  {\doibase 10.1088/1475-7516/2011/07/034} {\bibfield  {journal} {\bibinfo
  {journal} {JCAP}\ }\textbf {\bibinfo {volume} {1107}},\ \bibinfo {pages}
  {034} (\bibinfo {year} {2011})},\ \Eprint {http://arxiv.org/abs/1104.2933}
  {arXiv:1104.2933 [astro-ph.CO]} \BibitemShut {NoStop}%
%%CITATION = ARXIV:1104.2933;%%
\bibitem [{\citenamefont {Ir\v{s}i\v{c}}\ \emph {et~al.}(2017)\citenamefont
  {Ir\v{s}i\v{c}} \emph {et~al.}}]{Irsic:2017ixq}%
  \BibitemOpen
  \bibfield  {author} {\bibinfo {author} {\bibfnamefont {V.}~\bibnamefont
  {Ir\v{s}i\v{c}}} \emph {et~al.},\ }\href {\doibase
  10.1103/PhysRevD.96.023522} {\bibfield  {journal} {\bibinfo  {journal} {Phys.
  Rev. D}\ }\textbf {\bibinfo {volume} {96}},\ \bibinfo {pages} {023522}
  (\bibinfo {year} {2017})},\ \Eprint {http://arxiv.org/abs/1702.01764}
  {arXiv:1702.01764 [astro-ph.CO]} \BibitemShut {NoStop}%
%%CITATION = ARXIV:1702.01764;%%
\bibitem [{\citenamefont {Ade}\ \emph {et~al.}(2016)\citenamefont {Ade} \emph
  {et~al.}}]{Ade:2015xua}%
  \BibitemOpen
  \bibfield  {author} {\bibinfo {author} {\bibfnamefont {P.~A.~R.}\
  \bibnamefont {Ade}} \emph {et~al.} (\bibinfo {collaboration} {Planck}),\
  }\href {\doibase 10.1051/0004-6361/201525830} {\bibfield  {journal} {\bibinfo
   {journal} {Astron. Astrophys.}\ }\textbf {\bibinfo {volume} {594}},\
  \bibinfo {pages} {A13} (\bibinfo {year} {2016})},\ \Eprint
  {http://arxiv.org/abs/1502.01589} {arXiv:1502.01589 [astro-ph.CO]}
  \BibitemShut {NoStop}%
%%CITATION = ARXIV:1502.01589;%%
\bibitem [{\citenamefont {Seljak}\ \emph {et~al.}(2006)\citenamefont {Seljak},
  \citenamefont {Slosar},\ and\ \citenamefont {McDonald}}]{seljak2006}%
  \BibitemOpen
  \bibfield  {author} {\bibinfo {author} {\bibfnamefont {U.}~\bibnamefont
  {Seljak}}, \bibinfo {author} {\bibfnamefont {A.}~\bibnamefont {Slosar}}, \
  and\ \bibinfo {author} {\bibfnamefont {P.}~\bibnamefont {McDonald}},\ }\href
  {\doibase 10.1088/1475-7516/2006/10/014} {\bibfield  {journal} {\bibinfo
  {journal} {JCAP}\ }\textbf {\bibinfo {volume} {0610}},\ \bibinfo {pages}
  {014} (\bibinfo {year} {2006})},\ \Eprint
  {http://arxiv.org/abs/astro-ph/0604335} {arXiv:astro-ph/0604335 [astro-ph]}
  \BibitemShut {NoStop}%
%%CITATION = ASTRO-PH/0604335;%%
\bibitem [{\citenamefont {McDonald}\ \emph {et~al.}(2006)\citenamefont
  {McDonald} \emph {et~al.}}]{McDonald:2004eu}%
  \BibitemOpen
  \bibfield  {author} {\bibinfo {author} {\bibfnamefont {P.}~\bibnamefont
  {McDonald}} \emph {et~al.} (\bibinfo {collaboration} {SDSS}),\ }\href
  {\doibase 10.1086/444361} {\bibfield  {journal} {\bibinfo  {journal}
  {Astrophys. J. Suppl.}\ }\textbf {\bibinfo {volume} {163}},\ \bibinfo {pages}
  {80} (\bibinfo {year} {2006})},\ \Eprint
  {http://arxiv.org/abs/astro-ph/0405013} {arXiv:astro-ph/0405013 [astro-ph]}
  \BibitemShut {NoStop}%
%%CITATION = ASTRO-PH/0405013;%%
\bibitem [{\citenamefont {Arinyo-i Prats}\ \emph {et~al.}(2015)\citenamefont
  {Arinyo-i Prats}, \citenamefont {Miralda-Escude}, \citenamefont {Viel},\ and\
  \citenamefont {Cen}}]{Arinyo-i-Prats:2015vqa}%
  \BibitemOpen
  \bibfield  {author} {\bibinfo {author} {\bibfnamefont {A.}~\bibnamefont
  {Arinyo-i Prats}}, \bibinfo {author} {\bibfnamefont {J.}~\bibnamefont
  {Miralda-Escude}}, \bibinfo {author} {\bibfnamefont {M.}~\bibnamefont
  {Viel}}, \ and\ \bibinfo {author} {\bibfnamefont {R.}~\bibnamefont {Cen}},\
  }\href {\doibase 10.1088/1475-7516/2015/12/017} {\bibfield  {journal}
  {\bibinfo  {journal} {JCAP}\ }\textbf {\bibinfo {volume} {1512}},\ \bibinfo
  {pages} {017} (\bibinfo {year} {2015})},\ \Eprint
  {http://arxiv.org/abs/1506.04519} {arXiv:1506.04519 [astro-ph.CO]}
  \BibitemShut {NoStop}%
%%CITATION = ARXIV:1506.04519;%%
\bibitem [{\citenamefont {Hui}\ and\ \citenamefont {Gnedin}(1997)}]{hui97}%
  \BibitemOpen
  \bibfield  {author} {\bibinfo {author} {\bibfnamefont {L.}~\bibnamefont
  {Hui}}\ and\ \bibinfo {author} {\bibfnamefont {N.~Y.}\ \bibnamefont
  {Gnedin}},\ }\href {\doibase 10.1093/mnras/292.1.27} {\bibfield  {journal}
  {\bibinfo  {journal} {MNRAS}\ }\textbf {\bibinfo {volume} {292}},\ \bibinfo
  {pages} {27} (\bibinfo {year} {1997})},\ \Eprint
  {http://arxiv.org/abs/astro-ph/9612232} {arXiv:astro-ph/9612232 [astro-ph]}
  \BibitemShut {NoStop}%
%%CITATION = ASTRO-PH/9612232;%%
\bibitem [{\citenamefont {Bolton}\ \emph {et~al.}(2017)\citenamefont {Bolton},
  \citenamefont {Puchwein}, \citenamefont {Sijacki}, \citenamefont {Haehnelt},
  \citenamefont {Kim}, \citenamefont {Meiksin}, \citenamefont {Regan},\ and\
  \citenamefont {Viel}}]{bolton17}%
  \BibitemOpen
  \bibfield  {author} {\bibinfo {author} {\bibfnamefont {J.~S.}\ \bibnamefont
  {Bolton}}, \bibinfo {author} {\bibfnamefont {E.}~\bibnamefont {Puchwein}},
  \bibinfo {author} {\bibfnamefont {D.}~\bibnamefont {Sijacki}}, \bibinfo
  {author} {\bibfnamefont {M.~G.}\ \bibnamefont {Haehnelt}}, \bibinfo {author}
  {\bibfnamefont {T.-S.}\ \bibnamefont {Kim}}, \bibinfo {author} {\bibfnamefont
  {A.}~\bibnamefont {Meiksin}}, \bibinfo {author} {\bibfnamefont {J.~A.}\
  \bibnamefont {Regan}}, \ and\ \bibinfo {author} {\bibfnamefont
  {M.}~\bibnamefont {Viel}},\ }\href {\doibase 10.1093/mnras/stw2397}
  {\bibfield  {journal} {\bibinfo  {journal} {MNRAS}\ }\textbf {\bibinfo
  {volume} {464}},\ \bibinfo {pages} {897} (\bibinfo {year} {2017})},\ \Eprint
  {http://arxiv.org/abs/1605.03462} {arXiv:1605.03462 [astro-ph.CO]}
  \BibitemShut {NoStop}%
%%CITATION = ARXIV:1605.03462;%%
\bibitem [{\citenamefont {Palanque-Delabrouille}\ \emph
  {et~al.}(2013)\citenamefont {Palanque-Delabrouille} \emph
  {et~al.}}]{boss2013}%
  \BibitemOpen
  \bibfield  {author} {\bibinfo {author} {\bibfnamefont {N.}~\bibnamefont
  {Palanque-Delabrouille}} \emph {et~al.},\ }\href {\doibase
  10.1051/0004-6361/201322130} {\bibfield  {journal} {\bibinfo  {journal}
  {A\&A}\ }\textbf {\bibinfo {volume} {559}},\ \bibinfo {pages} {A85} (\bibinfo
  {year} {2013})}\BibitemShut {NoStop}%
\bibitem [{\citenamefont {Webster}\ and\ \citenamefont
  {Oliver}(2007)}]{webster2007geostatistics}%
  \BibitemOpen
  \bibfield  {author} {\bibinfo {author} {\bibfnamefont {R.}~\bibnamefont
  {Webster}}\ and\ \bibinfo {author} {\bibfnamefont {M.~A.}\ \bibnamefont
  {Oliver}},\ }\href@noop {} {\emph {\bibinfo {title} {Geostatistics for
  environmental scientists}}}\ (\bibinfo  {publisher} {John Wiley \& Sons},\
  \bibinfo {year} {2007})\BibitemShut {NoStop}%
\bibitem [{\citenamefont {{Foreman-Mackey}}\ \emph {et~al.}(2013)\citenamefont
  {{Foreman-Mackey}}, \citenamefont {{Hogg}}, \citenamefont {{Lang}},\ and\
  \citenamefont {{Goodman}}}]{emcee13}%
  \BibitemOpen
  \bibfield  {author} {\bibinfo {author} {\bibfnamefont {D.}~\bibnamefont
  {{Foreman-Mackey}}}, \bibinfo {author} {\bibfnamefont {D.~W.}\ \bibnamefont
  {{Hogg}}}, \bibinfo {author} {\bibfnamefont {D.}~\bibnamefont {{Lang}}}, \
  and\ \bibinfo {author} {\bibfnamefont {J.}~\bibnamefont {{Goodman}}},\ }\href
  {\doibase 10.1086/670067} {\bibfield  {journal} {\bibinfo  {journal}
  {{PASP}}\ }\textbf {\bibinfo {volume} {125}},\ \bibinfo {pages} {306}
  (\bibinfo {year} {2013})},\ \Eprint {http://arxiv.org/abs/1202.3665}
  {arXiv:1202.3665 [astro-ph.IM]} \BibitemShut {NoStop}%
\bibitem [{\citenamefont {Boera}\ \emph {et~al.}(2018)\citenamefont {Boera},
  \citenamefont {Becker}, \citenamefont {Bolton},\ and\ \citenamefont
  {Nasir}}]{Boera:2018vzq}%
  \BibitemOpen
  \bibfield  {author} {\bibinfo {author} {\bibfnamefont {E.}~\bibnamefont
  {Boera}}, \bibinfo {author} {\bibfnamefont {G.~D.}\ \bibnamefont {Becker}},
  \bibinfo {author} {\bibfnamefont {J.~S.}\ \bibnamefont {Bolton}}, \ and\
  \bibinfo {author} {\bibfnamefont {F.}~\bibnamefont {Nasir}},\ }\href@noop {}
  {\bibfield  {journal} {\bibinfo  {journal} {1809.06980}\ } (\bibinfo {year}
  {2018})}\BibitemShut {NoStop}%
%%CITATION = ARXIV:1809.06980;%%
\bibitem [{\citenamefont {Bartolo}\ \emph {et~al.}(2018)\citenamefont
  {Bartolo}, \citenamefont {De~Luca}, \citenamefont {Franciolini},
  \citenamefont {Peloso}, \citenamefont {Racco},\ and\ \citenamefont
  {Riotto}}]{Bartolo:2018rku}%
  \BibitemOpen
  \bibfield  {author} {\bibinfo {author} {\bibfnamefont {N.}~\bibnamefont
  {Bartolo}}, \bibinfo {author} {\bibfnamefont {V.}~\bibnamefont {De~Luca}},
  \bibinfo {author} {\bibfnamefont {G.}~\bibnamefont {Franciolini}}, \bibinfo
  {author} {\bibfnamefont {M.}~\bibnamefont {Peloso}}, \bibinfo {author}
  {\bibfnamefont {D.}~\bibnamefont {Racco}}, \ and\ \bibinfo {author}
  {\bibfnamefont {A.}~\bibnamefont {Riotto}},\ }\href@noop {} {\  (\bibinfo
  {year} {2018})},\ \Eprint {http://arxiv.org/abs/1810.12224} {arXiv:1810.12224
  [astro-ph.CO]} \BibitemShut {NoStop}%
%%CITATION = ARXIV:1810.12224;%%
\bibitem [{\citenamefont {Carr}\ \emph {et~al.}(2010)\citenamefont {Carr},
  \citenamefont {Kohri}, \citenamefont {Sendouda},\ and\ \citenamefont
  {Yokoyama}}]{Carr:2009jm}%
  \BibitemOpen
  \bibfield  {author} {\bibinfo {author} {\bibfnamefont {B.~J.}\ \bibnamefont
  {Carr}}, \bibinfo {author} {\bibfnamefont {K.}~\bibnamefont {Kohri}},
  \bibinfo {author} {\bibfnamefont {Y.}~\bibnamefont {Sendouda}}, \ and\
  \bibinfo {author} {\bibfnamefont {J.}~\bibnamefont {Yokoyama}},\ }\href
  {\doibase 10.1103/PhysRevD.81.104019} {\bibfield  {journal} {\bibinfo
  {journal} {Phys. Rev.}\ }\textbf {\bibinfo {volume} {D81}},\ \bibinfo {pages}
  {104019} (\bibinfo {year} {2010})},\ \Eprint {http://arxiv.org/abs/0912.5297}
  {arXiv:0912.5297 [astro-ph.CO]} \BibitemShut {NoStop}%
%%CITATION = ARXIV:0912.5297;%%
\bibitem [{\citenamefont {Carr}\ \emph {et~al.}(2016)\citenamefont {Carr},
  \citenamefont {K\"uhnel},\ and\ \citenamefont {Sandstad}}]{Carr:2016}%
  \BibitemOpen
  \bibfield  {author} {\bibinfo {author} {\bibfnamefont {B.}~\bibnamefont
  {Carr}}, \bibinfo {author} {\bibfnamefont {F.}~\bibnamefont {K\"uhnel}}, \
  and\ \bibinfo {author} {\bibfnamefont {M.}~\bibnamefont {Sandstad}},\ }\href
  {\doibase 10.1103/PhysRevD.94.083504} {\bibfield  {journal} {\bibinfo
  {journal} {Phys. Rev. D}\ }\textbf {\bibinfo {volume} {94}},\ \bibinfo
  {pages} {083504} (\bibinfo {year} {2016})}\BibitemShut {NoStop}%
\bibitem [{\citenamefont {Li}\ \emph {et~al.}(2017)\citenamefont {Li} \emph
  {et~al.}}]{Li:2016}%
  \BibitemOpen
  \bibfield  {author} {\bibinfo {author} {\bibfnamefont {T.~S.}\ \bibnamefont
  {Li}} \emph {et~al.} (\bibinfo {collaboration} {DES}),\ }\href {\doibase
  10.3847/1538-4357/aa6113} {\bibfield  {journal} {\bibinfo  {journal}
  {Astrophys. J.}\ }\textbf {\bibinfo {volume} {838}},\ \bibinfo {pages} {8}
  (\bibinfo {year} {2017})},\ \Eprint {http://arxiv.org/abs/1611.05052}
  {arXiv:1611.05052 [astro-ph.GA]} \BibitemShut {NoStop}%
%%CITATION = ARXIV:1611.05052;%%
\bibitem [{\citenamefont {Pi}\ \emph {et~al.}(2018)\citenamefont {Pi},
  \citenamefont {Zhang}, \citenamefont {Huang},\ and\ \citenamefont
  {Sasaki}}]{Pi:2017}%
  \BibitemOpen
  \bibfield  {author} {\bibinfo {author} {\bibfnamefont {S.}~\bibnamefont
  {Pi}}, \bibinfo {author} {\bibfnamefont {Y.-l.}\ \bibnamefont {Zhang}},
  \bibinfo {author} {\bibfnamefont {Q.-G.}\ \bibnamefont {Huang}}, \ and\
  \bibinfo {author} {\bibfnamefont {M.}~\bibnamefont {Sasaki}},\ }\href
  {\doibase 10.1088/1475-7516/2018/05/042} {\bibfield  {journal} {\bibinfo
  {journal} {JCAP}\ }\textbf {\bibinfo {volume} {1805}},\ \bibinfo {pages}
  {042} (\bibinfo {year} {2018})},\ \Eprint {http://arxiv.org/abs/1712.09896}
  {arXiv:1712.09896 [astro-ph.CO]} \BibitemShut {NoStop}%
%%CITATION = ARXIV:1712.09896;%%
\bibitem [{\citenamefont {Hawking}(1974)}]{hawking:1974radiation}%
  \BibitemOpen
  \bibfield  {author} {\bibinfo {author} {\bibfnamefont {S.~W.}\ \bibnamefont
  {Hawking}},\ }\href@noop {} {\bibfield  {journal} {\bibinfo  {journal}
  {Nature}\ }\textbf {\bibinfo {volume} {248}},\ \bibinfo {pages} {30}
  (\bibinfo {year} {1974})}\BibitemShut {NoStop}%
\bibitem [{\citenamefont {{Pepe}}\ \emph {et~al.}(2013)\citenamefont {{Pepe}}
  \emph {et~al.}}]{pepe13}%
  \BibitemOpen
  \bibfield  {author} {\bibinfo {author} {\bibfnamefont {F.}~\bibnamefont
  {{Pepe}}} \emph {et~al.},\ }\href@noop {} {\bibfield  {journal} {\bibinfo
  {journal} {The Messenger}\ }\textbf {\bibinfo {volume} {153}},\ \bibinfo
  {pages} {6} (\bibinfo {year} {2013})}\BibitemShut {NoStop}%
\bibitem [{\citenamefont {Kashlinsky}(2016)}]{Kashlinsky:2016sdv}%
  \BibitemOpen
  \bibfield  {author} {\bibinfo {author} {\bibfnamefont {A.}~\bibnamefont
  {Kashlinsky}},\ }\href {\doibase 10.3847/2041-8205/823/2/L25} {\bibfield
  {journal} {\bibinfo  {journal} {Astrophys. J.}\ }\textbf {\bibinfo {volume}
  {823}},\ \bibinfo {pages} {L25} (\bibinfo {year} {2016})},\ \Eprint
  {http://arxiv.org/abs/1605.04023} {arXiv:1605.04023 [astro-ph.CO]}
  \BibitemShut {NoStop}%
%%CITATION = ARXIV:1605.04023;%%
\bibitem [{\citenamefont {Kashlinsky}(2005)}]{Kashlinsky:2004jt}%
  \BibitemOpen
  \bibfield  {author} {\bibinfo {author} {\bibfnamefont {A.}~\bibnamefont
  {Kashlinsky}},\ }\href {\doibase 10.1016/j.physrep.2004.12.005} {\bibfield
  {journal} {\bibinfo  {journal} {Phys. Rept.}\ }\textbf {\bibinfo {volume}
  {409}},\ \bibinfo {pages} {361} (\bibinfo {year} {2005})},\ \Eprint
  {http://arxiv.org/abs/astro-ph/0412235} {arXiv:astro-ph/0412235 [astro-ph]}
  \BibitemShut {NoStop}%
%%CITATION = ASTRO-PH/0412235;%%
\bibitem [{\citenamefont {Kashlinsky}\ \emph {et~al.}(2005)\citenamefont
  {Kashlinsky}, \citenamefont {Arendt}, \citenamefont {Mather},\ and\
  \citenamefont {Moseley}}]{Kashlinsky:2005di}%
  \BibitemOpen
  \bibfield  {author} {\bibinfo {author} {\bibfnamefont {A.}~\bibnamefont
  {Kashlinsky}}, \bibinfo {author} {\bibfnamefont {R.~G.}\ \bibnamefont
  {Arendt}}, \bibinfo {author} {\bibfnamefont {J.~C.}\ \bibnamefont {Mather}},
  \ and\ \bibinfo {author} {\bibfnamefont {S.~H.}\ \bibnamefont {Moseley}},\
  }\href {\doibase 10.1038/nature04143} {\bibfield  {journal} {\bibinfo
  {journal} {Nature}\ }\textbf {\bibinfo {volume} {438}},\ \bibinfo {pages}
  {45} (\bibinfo {year} {2005})},\ \Eprint
  {http://arxiv.org/abs/astro-ph/0511105} {arXiv:astro-ph/0511105 [astro-ph]}
  \BibitemShut {NoStop}%
%%CITATION = ASTRO-PH/0511105;%%
\end{thebibliography}%

%%%%%%%%%%%%%%%%%%%%%%%%%%%%%%%%%%%%%%%%%%%%%%%%%%

%%%%%%%%%%%%%%%%% APPENDICES %%%%%%%%%%%%%%%%%%%%%

%\appendix


%%%%%%%%%%%%%%%%%%%%%%%%%%%%%%%%%%%%%%%%%%%%%%%%%%

\end{document}